\documentclass[aip,jmp,amsmath,amssymb,reprint]{revtex4-1}

\usepackage{graphicx}
\usepackage{amsmath}
\usepackage{amsfonts}
\usepackage{times}

\newcommand{\eg}[1]
  {{\it e.g.\/}\ifx#1.\else\expandafter#1\fi}
\newcommand{\eq}[1]{(\ref{#1})}         
\newcommand{\Fig}[1]{Figure~\ref{#1}}   
\newcommand{\Figs}[1]{Figures~\ref{#1}} 
\newcommand{\fig}[1]{Fig.~\ref{#1}}     
\newcommand{\figs}[2]{Figs.~\ref{#1}-\ref{#2}}  
\newcommand{\FIG}[2]{Fig.~\ref{#1}({#2})} 
\newcommand{\FIGG}[3]{Figs.~\ref{#1}({#2}) and \ref{#1}({#3})} 
\newcommand{\ie}[1]
  {{\it i.e.\/}\ifx#1.\else\expandafter#1\fi}
\newcommand{\secn}[1]{Sec.~\ref{sec:#1}} 
\newcommand{\seclabel}[1]{\label{sec:#1}}  

\newcommand{\dblfigure}[3]
  {\begin{figure*}[tbp]\includegraphics[width=1.0\textwidth]{#1.pdf}\caption[]{#2}\label{#3}\end{figure*}}
  \newcommand{\dblbotfigure}[3]
  {\begin{figure*}[b]\includegraphics[width=1.0\textwidth]{#1.pdf}\caption[]{#2}\label{#3}\end{figure*}}
\newcommand{\sglfigure}[3]
  {\begin{figure}[tbp]\centerline{\includegraphics[width=0.5\textwidth]{#1.pdf}}\caption[]{#2}\label{#3}\end{figure}}
\newcommand{\sglscfigure}[4]
  {\begin{figure}[tbp]\centerline{\includegraphics[width=#1]{#2.pdf}}\caption[]{#3}\label{#4}\end{figure}}

\renewcommand{\@}{\partial}             
  \newcommand{\<}{\langle}              
\renewcommand{\>}{\rangle}              
\newcommand{\crit}{_\mathrm{crit}}      
\newcommand{\conv}{_{\mathrm{conv}}}    
\newcommand{\inner}[2]
  {\left\<#1\, , \,#2\right\>}
\newcommand{\Mx}[1]{
\left[\begin{array}{cccccccc}#1\end{array}\right]}
\newcommand{\mx}[1]{\mathbf{#1}}        
\newcommand{\T}{^{\mathrm{T}}}          

\newcommand{\betzero}{\parbet_*}
\newcommand{\betone}{\parbet_\ell}
\newcommand{\bettwo}{\parbet_1}
\newcommand{\betthree}{\parbet_2}
\newcommand{\betfour}{\parbet_u}
\newcommand{\btwo}{b_2}
\newcommand{\cthree}{c_3}
\newcommand{\curv}{\kappa}              
\newcommand{\D}{\mx{D}}                 
\newcommand{\paralp}{\alpha}            
\newcommand{\parbet}{\beta}             
\newcommand{\pargam}{\gamma}            
\newcommand{\RF}[1]{\mx{W}^{(#1)}}      
\newcommand{\shockamp}{A}               
\newcommand{\h}{h}                      
\newcommand{\Tr}[1]{\mx{V}\ifx#1.\else^{(#1)}\fi} 
\newcommand{\U}{\mx{U}}                 
\newcommand{\eleamp}{\mathcal E}        

\newcommand{\dr}{\Delta r}              
\newcommand{\dt}{\Delta t}              
\newcommand{\dx}{\Delta x}              
\newcommand{\dy}{\Delta y}              
\newcommand{\dz}{\Delta z}              
\newcommand{\Nr}{N_{\rho}}              
\newcommand{\Nt}{N_{\theta}}            
\newcommand{\rp}{\rho_{\max}}           

\begin{document}

\title{Alternative Stable Scroll Waves and Conversion of Autowave Turbulence}

\author{A.~J.~Foulkes}
\affiliation{Department of Computer Science, University of Liverpool,
        Ashton Building, Ashton Street, Liverpool, L69 3BX, United Kingdom}
\author{D.~Barkley}
\affiliation{Mathematics Institute, University of Warwick, Coventry CV4 7AL, United Kingdom}
\author{V.~N.~Biktashev}
\affiliation{Department of Mathematical Sciences, University of Liverpool,
   Mathematical Sciences Building, Peach Street, Liverpool, L69 7ZL, United Kingdom}
\author{I.~V.~Biktasheva}
\affiliation{Department of Computer Science, University of Liverpool,
        Ashton Building, Ashton Street, Liverpool, L69 3BX, United Kingdom}

\date{\today}

\begin{abstract}
  Rotating spiral and scroll waves (vortices) are investigated in the
  FitzHugh-Nagumo model of excitable media. The focus is on a parameter
  region in which there exists bistability between alternative stable
  vortices with distinct periods.  \emph{Response Functions} are used to
  predict the filament tension of the alternative scrolls and it is
  shown that the slow-period scroll has negative filament tension, while
  the filament tension of the fast-period scroll changes sign within a
  hysteresis loop. The predictions are confirmed by direct
  simulations. Further investigations show that the slow-period scrolls
  display features similar to delayed after-depolarisation (DAD) and
  tend to develop into turbulence similar to Ventricular Fibrillation
  (VF). Scrolls with positive filament tension collapse or stabilize,
  similar to monomorphic Ventricular Tachycardia (VT).  Perturbations,
  such as boundary interaction or shock stimulus, can convert the vortex
  with negative filament tension into the vortex with positive filament
  tension.  This may correspond to transition from VF to VT unrelated to
  pinning.
\end{abstract}



\maketitle


\begin{quotation}
  Orderly contraction of the heart is essential to
  pump blood efficiently. This order is imposed by electrical
  excitation waves propagating throughout the heart muscle.
  Abnormalities in the propagation of excitation waves, known as
  arrhythmias, are responsible for many cardiac
  pathologies. Particularly dangerous are re-entrant arrhythmias, where
  excitation waves circulate along closed pathways.  Re-entrant
  excitation ``vortices'', that are not attached to anatomical
  features, are spiral waves (in two dimensions) scroll
  waves (in three dimensions). As a rule, the rotation frequency and
  shape of a spiral wave is uniquely determined by local properties
  of the tissue.  We investigate a region of parameters in a
  simplified mathematical model of excitation, in which two classes of
  vortices can exist. They differ significantly in their frequency
  and shape. Which of the two is realized, depends on the initial
  conditions.  The defining feature of the model, responsible for this
  dichotomy, is well known in heart electrophysiology. It is called
  delayed after-depolarization and is known to have a pro-arrhythmic effect. 
  We focus
  especially on the three-dimensional scroll waves.  Based on the
  asymptotic theory utlizing Response Functions, we predict a change
  of sign of the filament tension of the scrolls on one of the
  branches. Negative tension is known to promote ``scroll wave
  turbulence''. We investigate, by numerical simulations, specifics of
  this phenomenon in presence of alternative scrolls. We describe
  conversion of one type of vortex to the other under various 
  perturbations, including external shocks such as are used for
  defibrillation, interactions with boundaries, and
  curvature of the scroll filaments. Finally, we discuss implications
  of our findings for cardiac arrhythmias.
\end{quotation}



\section{Introduction}

Spiral waves in two-dimensions, and scroll waves in three-dimension, are
regimes of self-organization observed in physical
\cite{%
  Frisch-etal-1994,%
  Madore-Freedman-1987,%
  Schulman-Seiden-1986%
}, chemical \cite{%
  Zhabotinsky-Zaikin-1971,%
  Jakubith-etal-1990%
}, and biological \cite{%
  Allessie-etal-1973,%
  Gorelova-Bures-1983,%
  Alcantara-Monk-1974,%
  Lechleiter-etal-1991,%
  Carey-etal-1978,%
  Murray-etal-1986%
} 
dissipative systems, where wave propagation is supported by a source of energy
stored in the medium.
The interest in the dynamics of these waves has significantly broadened over
the years as developments in experimental techniques have permitted them to be
observed and studied in an ever increasing number of diverse
systems~\cite{%
  Shagalov-1997,%
  Oswald-Dequidt-2008,%
  Larionova-etal-2005,%
  Yu-etal-1999,%
  Agladze-Steinbock-2000,%
  Bretschneider-etal-2009,%
  Dahlem-Mueller-2003,
  Igoshin-etal-2004%
}. However, the occurrence of these waves in excitable media, and cardiac
tissue in particular, has been and continues to be one of the main motivating
factors for their study.
 
In most situations, a spiral wave in excitable media rotates with a period
determined uniquely by the medium. However, there are cases in which there is
bistability between alternative waves in the same medium. It is precisely this
case which is of interest in this paper. Our particular concerns are (1) the
transitions between the alternative solutions, that is, how can transition
from one type of solution to the other be effected, (2) the differences in the
dynamics exhibited by alternative scroll waves in the 3D setting, and finally
(3) the relationship these phenomena may have to cardiac electrophysiology.

Our study is based on the FitzHugh-Nagumo (FHN) model, a standard
two-component reaction-diffusion model capturing the essential features of
excitable media such as cardiac tissue. The model is given by
\begin{align}
\@_t u & = f(u,v) + \nabla^2 u, \label{FHN1} \\
\@_t v & = g(u,v),\label{FHN2} 
\end{align}
where $u$ and $v$ are the dependent variables, corresponding roughly to
membrane voltage and ionic channels, respectively. The kinetic terms in
the model are
\begin{align}
  f(u,v) & = \paralp^{-1}(u-u^3/3-v),                         \nonumber\\
  g(u,v) & = \paralp \, (  u + \parbet - \pargam v ).        \nonumber
\end{align}
with parameters $\paralp$, $\parbet$, and $\pargam$.
Space units are set such that the diffusion
coefficient is 1.

Typical solutions to the FHN model include rotating spiral waves in 2D and
rotating scroll waves in 3D.  These waves, whether in 2D or 3D, are commonly
referred to vortices, although they are unrelated to fluid vorticity.  
As already noted, in most situations vortices rotate with a period 
determined uniquely by the model parameters.  However, it has been shown by
Winfree~\cite{Winfree-1991a} that for certain model parameters there exist
alternative stable spirals with distinct periods.  A pair of such stable
spiral solutions in the FHN model is shown in \fig{intro_own}.
We refer to the two solutions as slow and fast.  The slow spiral has a
longer temporal period, larger spatial wavelength, and larger core
radius than the fast one. The longer period of a slow vortex is due to
the small extra loop near the fixed point in
the $u$-$v$ phase portrait, \FIG{intro_own}{e},
which manifests itself as extra maximum in the tail of the action
potential, \FIG{intro_own}{c}: the feature known as delayed after-depolarisation (DAD) in cardiac
electrophysiology~\cite{Wu-etal-2004}.
The fast, shorter period vortices do not show DADs
in their action potentials, \FIGG{intro_own}{d}{f}.

\sglfigure{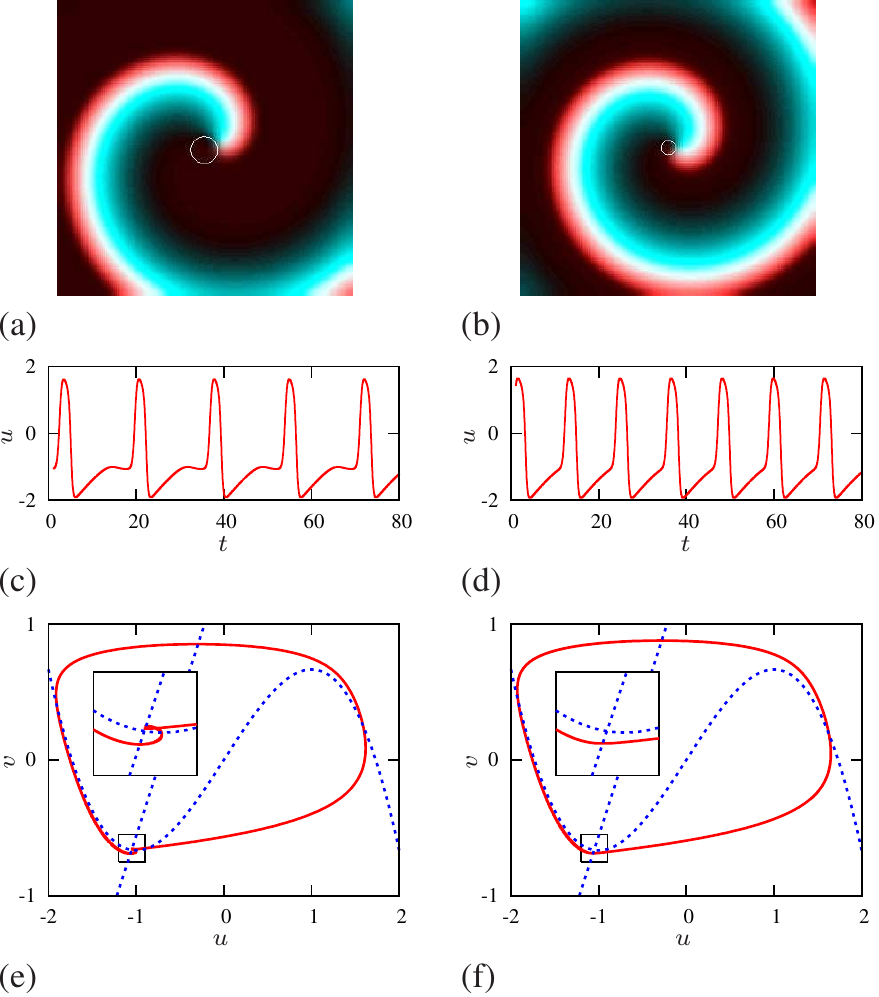}{
  (Color online) A pair of alternative vortices for the same 
  model parameters: $\paralp=0.3, \; \parbet=0.72, \;
  \pargam=0.5$. 
  (Left) slow solution and (right) fast solution. (a,b) spiral
  waves, shown is the $u$-field, and tip trace; (c,d) action potentials; (e,f) phase portraits,
  enlarged is the area around the fixed point.
}{intro_own}

The shaded region in \fig{parport} shows the portion of parameter space, for
fixed $\pargam=0.5$, where alternative vortices exist. This region has been
computed as part of the present study. Fast spirals exist to the lower left
while slow spirals exist to the upper right. Within the cusp-shaped region
both types of vortices exist with bistability between them. The boundaries of
the bistable region are fold singularies (limit points) meeting at a cusp,
$(\paralp,\parbet)\approx(0.27,0.8)$. Above the cusp, fast and slow vortices
are connected continuously. The thin vertical line shows a representative
one-parameter cut, $\paralp=0.3$, that will be the focus of much of our study.  The dashed
line relates to a 3D phenomenon which we now address.

\sglscfigure{0.35\textwidth}{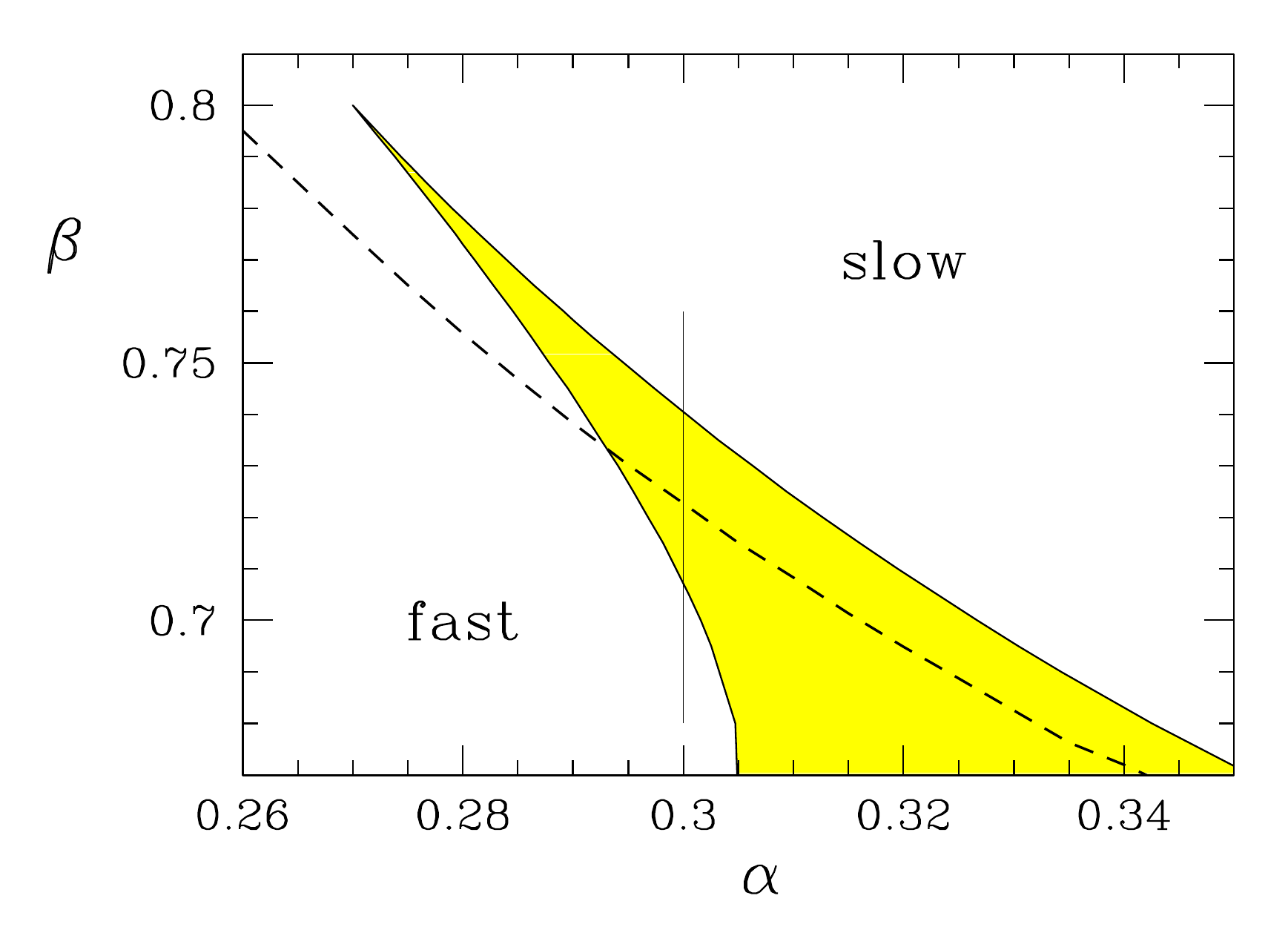}{
  (Color online) 
  Portion of the parameter space of the FitzHugh-Nagumo model containing
  alternative vortices. $\pargam$ is fixed at $0.5$.  Within the shaded
  cusp-shapped region both fast and slow vortices exist.  The thin vertical
  line at $\alpha=0.3$ shows the parameter cut considered throughout the
  paper. The dashed curve marks a line of zero filament tension.  To the
  lower left of this curve fast vortices have positive filament
  tension.
}{parport}

In 3D, scroll waves are organized about filaments and the possible behavior is
richer than in 2D. Filaments are not, in general, fixed in space but instead
undergo motion, typically on a slow timescale relative to the rotation
period. Hence, in addition to whatever dynamics 2D spiral waves might have,
scroll waves exhibit additional dynamics associated with filament
motion~\cite{%
  Yakushevich-1984,%
  Panfilov-Pertsov-1984,%
  Panfilov-etal-1986,%
  Panfilov-Rudenko-1987%
}. Working in Frenet coordinates, the motion may be conveniently expressed in
terms of the velocities $V_N$ and $V_B$ in the normal and binormal
directions, respectively, at each point along the filament. Motion along the
tangential direction is of no physical significance and is equivalent to
reparametrization of the filament.

First semi-phenomenologically~\cite{Brazhnik-etal-1987} and later using
asymptotics~\cite{Keener-1988,Biktashev-1989,Biktashev-etal-1994}, the
equations of filament motion have been obtained. At lowest order these are
\begin{equation}
 V_N = \btwo\curv + \dots,  \qquad
 V_B = \cthree\curv + \dots,  \label{fil-motion}
\end{equation}
where $\curv$ is the filament curvature. The coefficients $\btwo$ and
$\cthree$ depend on properties of the medium and 2D spiral solutions.
The coefficient $\btwo$ is called the {\em filament tension}. To understand
this, consider a circular filament, i.e.\ a scroll ring.  For positive $b_2$ a
scroll ring will contract somewhat as if the ring were an elastic ring under
tension.  For negative $b_2$, the ring will expand, as though under negative
tension. More generally in the negative-tension case, filaments will increase
in length and eventually evolve into full scale autowave
turbulence~\cite{Biktashev-1998}.
The dashed curve in \fig{parport} indicates a change of sign of filament
tension discussed at length later in the paper. 
The coefficient $c_3$, which we shall call the {\em binormal drift coefficient}
describes the drift of a scroll ring perpendicular to the plane of the ring,
or more generally, the velocity component orthogonal to the local plane of the
filament.

The remainder of the paper is then devoted to understanding the alternative
vortices in 2D and 3D. We first consider relevant asymptotic theory and
discuss how Response Functions can be used to predict the filament tension
of the alternative vortices. We then investigate the dynamics of the
alternative vortices in 2D and 3D through direct numerical simulations.  In
addition to confirming the prediction, we study how the alternative vortices
can be converted one into another in a variety of situations.



\sglscfigure{0.35\textwidth}{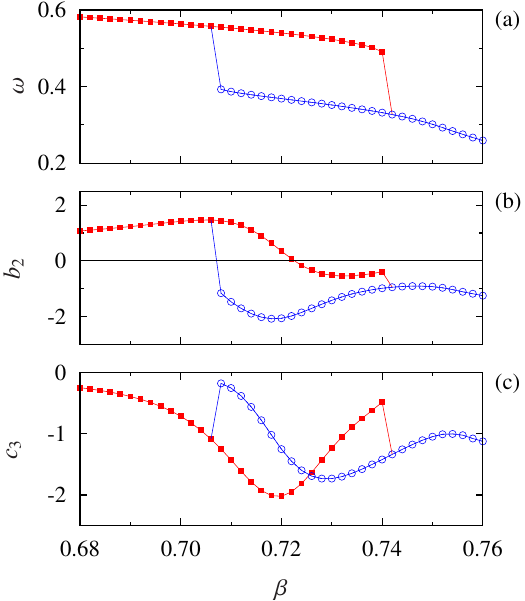}{
  (Color online) Alternative vortices along the parameter cut
  $\paralp=0.30,\: \pargam=0.5$ shown in \fig{parport}. Red squares are
  fast spirals and blue circle are slow spirals. (a) Hysteresis loop in
  the rotation velocity $\omega$. (b) Prediction for filament tension
  $b_2$. (c) Prediction for the binormal drift coefficient $c_3$.
}{fig1}

\section{Asymptotical Predictions}
\seclabel{predict}


\subsection{Filament tension and drift}

At leading asymptotic order, the motion of scroll filaments is determined by
the two coefficients $\btwo$ and $\cthree$ appearing in
Eqs.~\eq{fil-motion}. Once these coefficients are known, the most important
properties of filament dynamics, especially the sign of the coefficient
$\btwo$ determining the sign of the filament tension, are determined.

The coefficients for filament motion in Eqs.~\eq{fil-motion} are given by
the following simple formula~\cite{Biktashev-1989,Biktashev-etal-1994}
\begin{equation}
\btwo+ i\cthree = \inner{ \RF{1} }{ \D\Tr{1} }, \label{b2_coeff}
\end{equation}
where angle brackets denote an inner product, $\Tr{1}$ is a Goldstone
mode, $\RF{1}$ is the corresponding Response Function (RF), and $\D$ is a
diffusion matrix. We now elaborate, although we refer the reader to other
publications for most technical details~\cite{%
  Keener-1988,%
  Biktashev-etal-1994,%
  Biktashev-Holden-1995,%
  Biktasheva-Biktashev-2003,%
  Biktasheva-etal-2010%
}.

First consider a rotating spiral wave solution of
Eqs.~\eq{FHN1}-\eq{FHN2}.  It is convenient to work in a frame of
reference co-rotating with the spiral, at angular velocity $\omega$. Denote the
steady spiral seen in this co-rotating frame by $\U=(u,v)\T$.

Now consider the linear
stability of such a spiral. Due to symmetries of the system, there will be
three eigenvalues given by $\lambda = i n \omega$, with corresponding
eigenvectors or Goldstone modes $\Tr{n}$, where $n=-1, 0, 1$. The $n=0$
eigenvalue and Goldstone mode is related to rotational symmetry and the
complex pair of eigenvalues and Goldstone modes $n=\pm 1$ is related to
translational symmetry.
 
In addition to the linear stability problem there is the associated adjoint
problem. The adjoints eigenmodes corresponding to
eigenvalues $-i n \omega$ are the RFs and are denoted $\RF{n}$. Thus one sees
that to obtain the coefficients for filament motion in Eq.~\eq{b2_coeff}
we require the translational Goldstone mode $\Tr{1}$ and corresponding
Response Function $\RF{1}$. $\D$ is just the matrix of diffusion
coefficients appearing in the reaction-diffusion
equations, so for Eqs.~\eq{FHN1}-\eq{FHN2}, 
$\mx{D}=\Mx{1&0\\0&0}$.  
The angle brackets in Eq.~\eq{b2_coeff} signify a straightforward
integration over space of the Hermitian product of the vector fields. 
We note that similar methods employing RFs can be used to obtain the
drift velocities of vortices in response to perturbations to
Eqs.~\eq{FHN1}-\eq{FHN2}~\cite{%
  Biktasheva-etal-1999,%
  Biktasheva-2000,%
  Henry-Hakim-2002,%
  Biktasheva-etal-2010%
}.

The prediction of filament tension, and more generally drift
velocities, relies on two conditions: firstly, the localization of the RFs
$\RF{n}$ in the vicinity of the core of the spiral and secondly, the ability
to compute the RFs efficiently and accurately.
Existence of localized RFs has been demonstrated 
for a broad range of the models' parameters in 
several models~\cite{%
  Hamm-1997,%
  Biktasheva-etal-1998,%
  Biktasheva-Biktashev-2001,%
  Henry-Hakim-2002,%
  Biktasheva-etal-2006,%
  Biktasheva-etal-2009,%
  Biktashev-etal-2010a%
}. 
A robust method to compute the RFs with good accuracy for any model of
excitable medium with differentiable kinetics has been developed
in Ref.~\onlinecite{Biktasheva-etal-2009}, 
extending eariler stability methods~\cite{Wheeler-Barkley-2006}.
Using these methods, we have computed 
the steady spiral $\U$, its rotational velocity $\omega$
together with the Goldstone modes $\Tr{n}$ and the response functions
$\RF{n}$ for model~\eq{FHN1}-\eq{FHN2}
on a disk of radius $\rp=25$ with $\Nt=64$ grid points
in the angular direction and $\Nr=251$ grid points in the radial
direction, see~\fig{rfs}.

\sglfigure{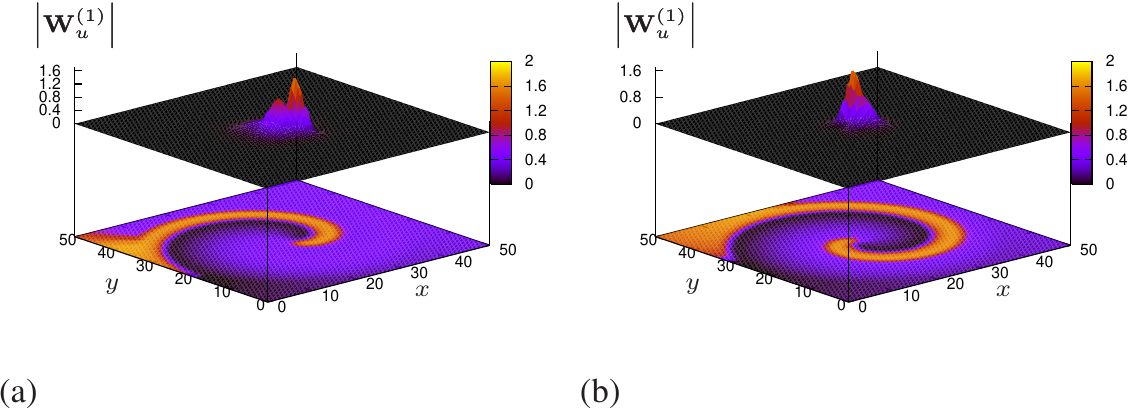}{
  (color online) Spiral waves, shown is the $u$-field (below, flat),
  and their translational response functions $\left|\RF{1}_u\right|$
  (above, elevated surface) for (a) the slow solution, and (b) the
  fast solution; $\parbet=0.72$.
}{rfs}

We focus now on the parameter path indicated in \fig{parport}.  For
$\paralp=0.3$ and $\pargam=0.5$ the hysteresis loop in the spirals' rotational
velocity $\omega$ has been obtained by continuation in parameter $\parbet$.
As is seen in \FIG{fig1}{a}, for these parameters, alternative stable spiral
wave solutions with distinct $\omega$ exist in the range
$\betone \le \parbet \le \betfour$, where $\betone\approx0.708$,
$\betfour\approx0.740$.

We have also computed the filament tension $b_2$ and the drift coefficient
$c_3$ along this parameter path and the results are shown in
\FIGG{fig1}{b}{c}. The most significant feature of these results
is the sign change of the filament tension within the hysteresis loop.  Below
$\parbet=\betzero \approx 0.722$, the alternative vortices have filament
tension of opposite signs: the fast vortex has positive tension while the slow
one has negative tension. At $\parbet=\betzero$, the filament tension of the
fast vortex changes sign so that in the parameter range
$\betzero \le \parbet \le \betfour$, the alternative vortices both have
negative filament tension.

The drift coefficient $c_3$ computed in this parameter cut is of fixed
sign (negative). The graphs of this drift coefficient of the
alternative vortices cross at $\parbet \approx 0.726$.


\subsection{Scroll rings and electrophoretic drift}
\seclabel{rings}

There is a connection between filament motion of scroll rings and the drift of
spiral waves in response to applied electric fields (electrophoretic
drift). Both things will appear in the subsequent numerical studies and it is
appropriate to summarize the issues here.

A scroll ring not only has a circular filament, but the entire
solution has an axial symmetry. 
It is convenient to study such a
structure in cylindrical coordinates $(r,\theta,z)$, where by symmetry, the
solution is independent of $\theta$. Hence 
the diffusion term in Eq.~\eq{FHN1} becomes
\begin{equation*}
\nabla^2 u(r,z) = \left ( \frac{\partial^2}{\partial r^2} + 
\frac{1}{r}\frac{\partial}{\partial r} + \frac{\partial^2}{\partial z^2} 
\right ) u(r,z).  
\end{equation*}
In this way scroll rings may be studied using 2D numerical computations as
long as one is not interested in any symmetry-breaking effects. 
Moreover, if one is interested in scroll rings with small curvature $\kappa$,
corresponding to rings of large filament radius $R = 1/\kappa$, one may use the
approximation
\begin{equation}
\nabla^2 u(r,z) \approx \left ( \frac{\partial^2}{\partial r^2} + 
\kappa \frac{\partial}{\partial r} + \frac{\partial^2}{\partial z^2}  
\right ) u(r,z),  
\label{cyl_lap}
\end{equation}
For small-curvature rings this is an exceedingly accurate approximation. In
fact it is equivalent to considering only the lowest-order curvature
contributions as in Eqs.~\eq{fil-motion}.  We shall use this approach to
evaluate filament tension and drift.  Note that the normal to the scroll ring
points in the outward radial direction and so a positive normal velocity $V_N>0$
corresponds to decreasing $r$, and vice versa.

Returning to the 2D Cartesian situation, if the excitable medium 
is a reaction-diffusion system in which the $u$ reagent is
electrically charged and the $v$ is neutral, then 
the effect of applying an electric
field can be modeled by including a term
\begin{equation}
\eleamp \frac{\partial{u}}{\partial{x}},             \label{electr_pert}
\end{equation}
on the right hand side of Eq.~\eq{FHN1}, 
where $\eleamp$ is proportional
to the applied electric field which is taken to be in the $x$
direction. 

For
small values of $\eleamp$, the effect of such a term will be drift of
the spiral and
one can calculate, using Response Functions, the drift
velocity~\cite{Henry-Hakim-2002,Biktasheva-etal-2010}.
However, one can use the formal
equivalence~\cite{Panfilov-etal-1986} between the term
$\eleamp \partial{u}/\partial{x}$ of Eq.~\eq{electr_pert} and
the term $\kappa \partial{u}/\partial r$ in Eq.~\eq{cyl_lap}
to obtain the drift velocity for vortex rings
from the results for the applied electric field.
Note, that electrophoretic drift towards negative $x$
corresponds to positive filament tension and towards positive $x$ corresponds
to negative filament tension.



\section{Two-Dimensional Simulations}
\seclabel{2D}

2D simulations have been performed with a suitably modified version of
EZSPIRAL~\cite{Barkley-1991,ezspiral}.
Unless specified otherwise, simulations have
been performed using forward Euler timestepping on a uniform Cartesian grid on
square domains $40\times40\:$s.u. with non-flux boundary conditions,
nine-point approximation of the Laplacian, spatial discretization 
$\dx=\dy=\h=1/3$ and time step $\dt=3/80$. The tip of the spiral is
defined as the intersections of isolines $u(x,y)=u_*$ and $v(x,y)=v_*$, and
the angle between $\nabla u$ at the tip and $x$ axis is taken as orientation
of the tip.  We use $(u_*,v_*)=(0,0)$ for the FitzHugh-Nagumo model.  Initial
conditions for the alternative spirals have been computed using parameter
continuation and then converted into the input format of EZSPIRAL.


\subsection{Alternative vortices: conversion by a shock}

First, to test if the alternative vortices can be converted from one into
another in a controllable way, we add a constant value $\shockamp$ to the
fast variable $u$, uniformily in space, at a specified time instant $T$. 
\begin{equation}
        u(x,y,t) \rightarrow u(x,y,t)+\shockamp, \quad \mbox{at $t=T$}
\label{shock}
\end{equation}
This may be considered as a very crude model of a defibrillating shock.
We use a box size of $50\times50\:$s.u., and in selected simulations up to
$150\times150\:$s.u., with spatial discretization $\h=0.1$ and time step
$\dt=2.25\times10^{-3}$.

\sglscfigure{0.5\textwidth}{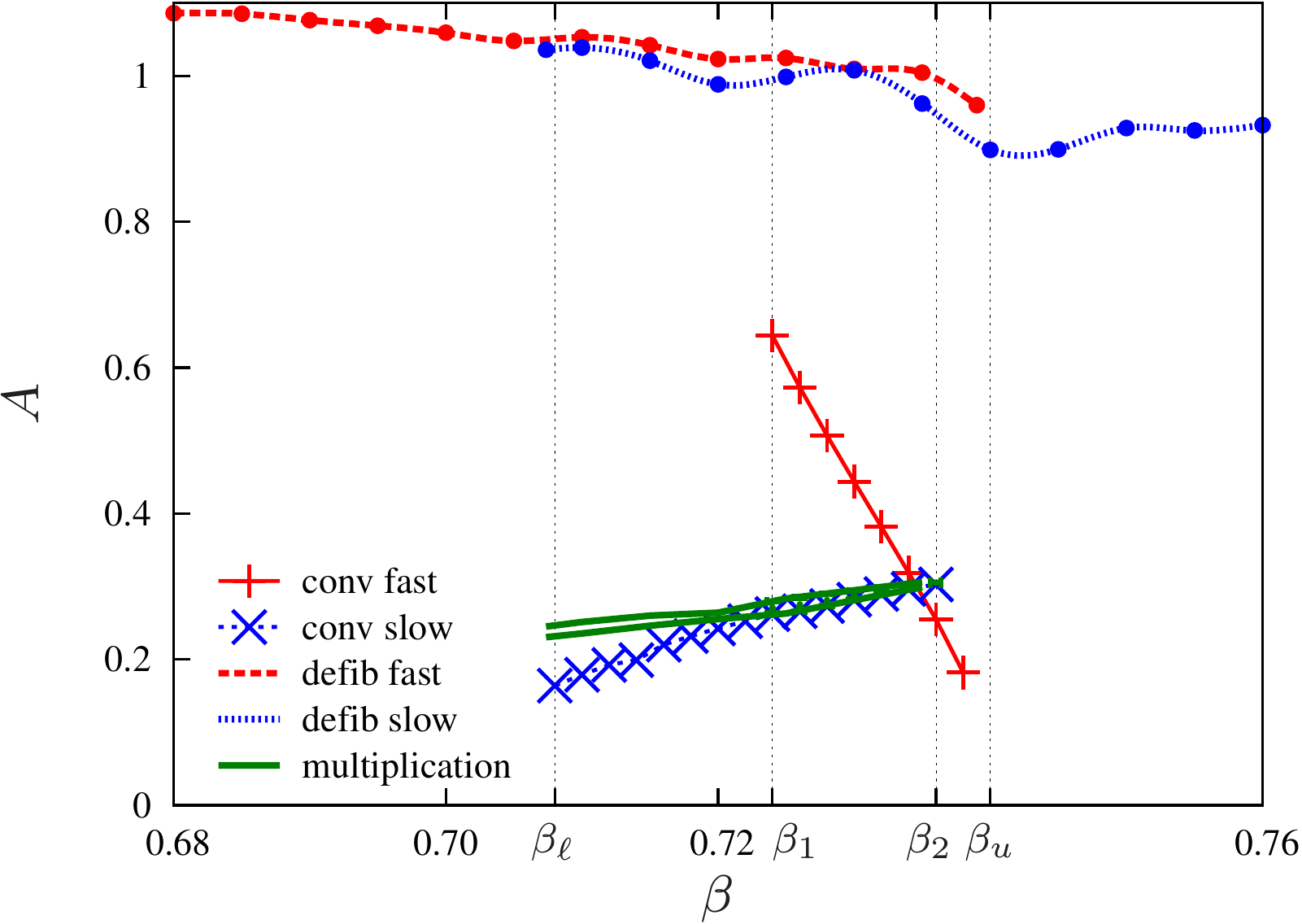}{
  (Color online) Minimum shock amplitudes $\shockamp$ needed to
  convert one type of spiral into the other (lines with large
  symbols), in comparison with the corresponding defibrillation
  thresholds (dashed and dotted lines). In the legend, 'conv
  fast'/'conv slow' refers to the shocks applied to fast/slow spirals
  necessary for conversion. Likewise, 'defib fast'/'defib slow' refers
  to the shocks applied to fast/slow spirals necessary for
  defibrillation.  The thick solid lines above the 'conv slow' curve
  indicate the region of parameters causing conversion with
  multiplication via front break-up.
}{b2}

The minimum shock amplitudes required to convert one alternative vortex into
another are shown in \fig{b2}, together with the corresponding
``defibrillation thresholds'' that are sufficient for the complete elimination
of spiral-wave activity.

It can be seen that, within the parameter range of the hysteresis
$\betone \le \beta \le \betfour$, there are three distinct intervals of
behavior demarked by the values $\bettwo\approx0.722$ and
$\betthree\approx0.736$.
In the region between $\betone$ and $\bettwo$ it is possible to convert a
slow vortex into its faster counterpart.  Transitions in the opposite
direction have not been observed, even as we increase $\shockamp$ to the
defibrillation threshold, which eliminates spirals from the medium.
In the middle interval, between $\bettwo$ and $\betthree$, conversion both
ways is possible.
For $\parbet$ in the range between $\betthree$ and $\betfour$, we only
observe conversions from fast vortices into their slow counterparts, but not
vice versa.

The value of $\bettwo$, which is the smallest
$\beta$ for which conversion is possible from fast to slow
vortices with a single shock, is close to $\betzero$ at which the filament tension of
the fast vortices is zero.
This is a purely empirical observation and we have not investigated how
precisely it holds.

We also
observed \textit{conversion with multiplication} from a slow vortex with a large
core into multiple fast vortices with small cores, via break-up of some
segments of excitation fronts. This is illustrated in \fig{multi}.  This
conversion with multiplication occurs in a very small range of shock
amplitudes just above the conversion amplitude; the numerical values are
listed in Table~\ref{tab:multispiral}.  Note that break-up of spiral waves by
spatially uniform shocks has been previously
observed, \eg\ Ref.~\onlinecite{Keener-Panfilov-1996}, in cases without bistability, but
at shock magnitudes close to the defibrillation threshold. Here we observe it
in a system with alternative spiral waves, at shock magnitudes significantly
smaller than the defibrillation threshold and only slightly above the
conversion threshold.  We have not observed conversion with multiplication
from a fast spiral into multiple slow spirals by any shock amplitude.

\sglfigure{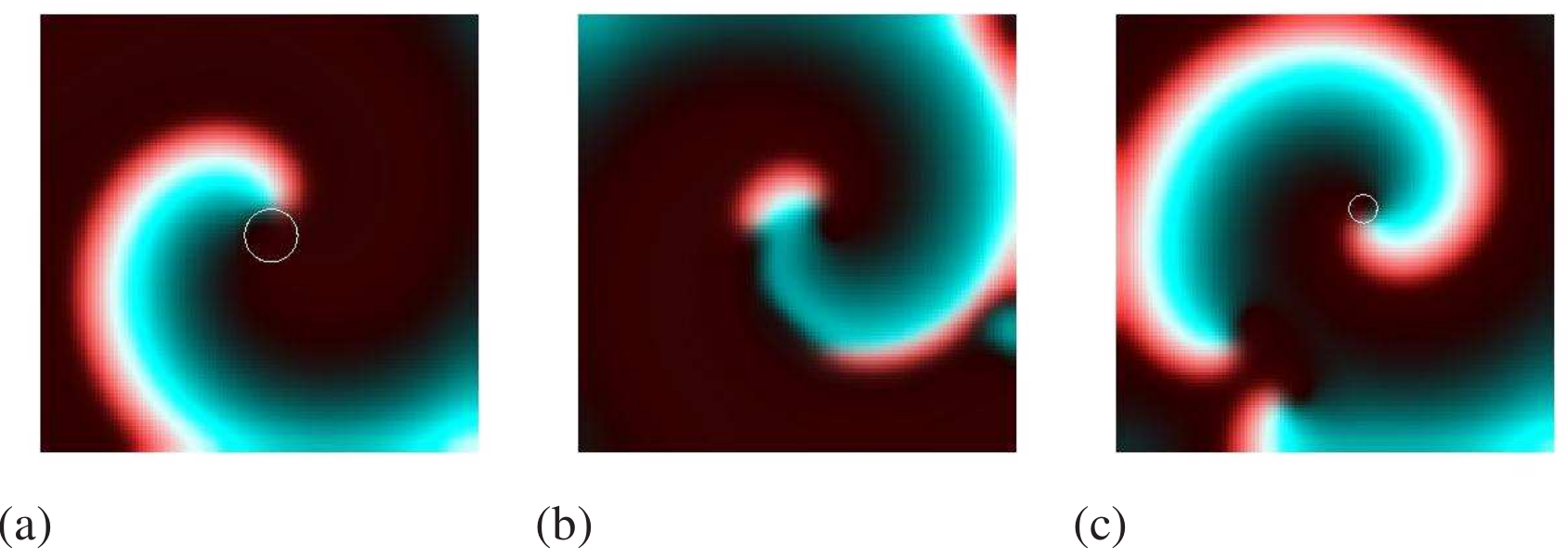}{
  (Color online) Shock conversion with
  multiplication at $\parbet=0.73$. 
  (a) Slow vortex with the large core before the shock. 
  (b) Front breaks immediately after a shock of amplitude $\shockamp=0.29$.
  (c) Multiple fast spirals with small cores, 
   approximately $50.3$ t.u. after the shock.
}{multi}

\begin{table}
\begin{ruledtabular}
\begin{tabular}{lll}
$\parbet$ & $\shockamp_{\min}$ & $\shockamp_{\max}$ \\
\hline \\
0.70735	& 0.2306 & 0.2452\\
0.710   & 0.2357 & 0.2511\\
0.715	& 0.2463 & 0.2602\\
0.720   & 0.2552 & 0.2642\\
0.724   & 0.2625 & 0.2805\\
0.725	& 0.2628 & 0.2834\\
0.726   & 0.2657 & 0.2850\\
0.728   & 0.2730 & 0.2896\\
0.730	& 0.2813 & 0.2946\\
0.730   & 0.2818 & 0.2941\\
0.732   & 0.2895 & 0.2986\\
0.734   & 0.2965 & 0.3032\\
0.735	& 0.2991 & 0.3059\\
0.736   & 0.3027 & 0.3077\\
\end{tabular}
\end{ruledtabular}
\caption{The minimum $\shockamp_{\min}$ and maximum $\shockamp_{\max}$ shock 
amplitudes producing multiple spirals from a slow vortex.}
\label{tab:multispiral}
\end{table}


\subsection{Conversion due to interaction with the boundary}
\seclabel{2Dconversion}
\dblfigure{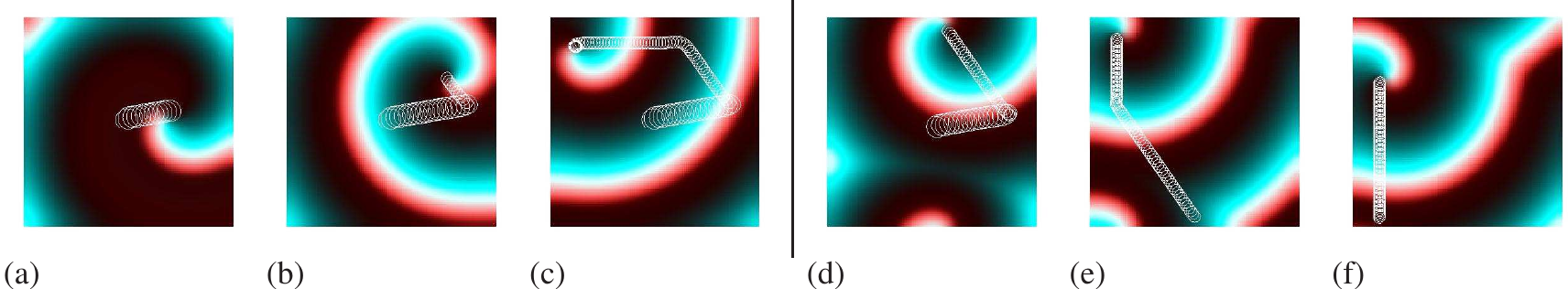}{
  (Color online) Conversion due to do interaction with a boundary.
  Simulations are started with a slow spiral at $\parbet=0.71$.  The spiral is
  induced to drift toward the right boundary by electrophoresis
  perturbation of amplitude $\eleamp=0.03$.
  (a-c) Three
  snapshots of the evolution with Neumann boundaries conditions on all four
  sides.  (d-f) Three snapshots of the evolution with Neumann boundaries
  conditions on right and left, and periodic boundary conditions 
  on top and bottom.
}{electro_nbc_0.71}

To study how interaction with a boundary may bring about conversion, we 
gently push the spiral towards a boundary using electrophoretic drift
discussed in \secn{rings}. 
Recall that negative tension ($\btwo<0$) corresponds to electrophoretic drift
to the right and positive tension ($\btwo>0$) corresponds to drift to the
left.

Representative simulations are shown in \fig{electro_nbc_0.71} at
$\parbet=0.71$.
Simulations are started from the slow, large-core spiral.  The period of the
spiral is $T\approx16.8654$.  The spiral drifts from its initial position to
the right in agreement with the predicted negative filament tension, $\btwo <
0$. It also moves up, as it should, since predicted drift coefficient,
$\cthree$ is negative.
When the spiral nears a boundary, its core radius decreases significantly and
its period changes to $T\approx11.2964$.  The horizontal drift component
changes sign [\FIG{electro_nbc_0.71}{b}], corresponding to switching to
positive filament tension, and the spiral moves to the left rather than to the
right. The binormal component of drift changes magnitude but not sign.

At the top boundary, the spiral drifts to the left until it pins to the top
left corner of the box [\FIG{electro_nbc_0.71}{c}].  If the Neumann boundary
conditions (NBCs) at the top and bottom edges of the box are changed to
periodic boundary conditions (PBCs), the fast spiral with a small core will
continue to drift upwards along the left boundary indefinitely
[\FIG{electro_nbc_0.71}{d-f}].


\subsection{Conversion due to applied field}
\seclabel{2Delectro}

\dblfigure{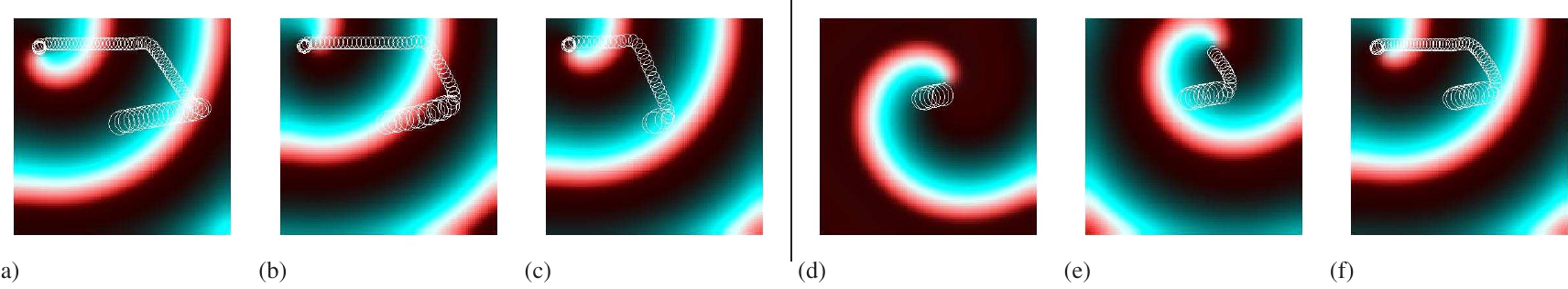}{
  (Color online) Conversion due to allied
  fields.  All simulations start from a slow spiral at $\parbet=0.71$. (a-c)
  Three simulations for field strengths (a) $\eleamp=0.035$, (b)
  $\eleamp=0.0472$, (c) $\eleamp=0.06$. In (a) conversion occurs at the
  boundary, whereas for (b,c) conversion occurs due to sufficiently strong
  electrophoretic drift.  (d-f) Three snapshots of a single simulation with
  $\eleamp=0.03$. After 8 rotations a shock of strength $\shockamp=0.14889$ is
  applied resulting in conversion.
}{electro_shock_amps}

Electrophoretic driving of a sufficiently high magnitude can itself cause
conversion of vortices directly. \Fig{electro_shock_amps} shows selected
simulations illustrating this phenomenon.
\Fig{electro_shock_amps}(a) shows that $\eleamp=0.035$ is not strong enough 
to convert the slow spiral into the fast, and indeed it allows the spiral to
continue to drift to the right boundary, at which the conversion
happens. \Fig{electro_shock_amps}(b) shows the simulation with the threshold
amplitude $\eleamp=0.0472$, allowing the conversion to happen well before
reaching the boundary. Finally, \FIG{electro_shock_amps}{c} shows a simulation
with $\eleamp=0.06$ which is strong enough to convert the
slow spiral into the fast one almost immediately.


\Figs{electro_shock_amps}(d-f) illustrate the combined effect
of an electrophoretic driving and defibrillation-style shock.  Shocks of
various amplitudes are applied after 8 rotation periods of the slow spiral. If
the amplitude is sufficiently high, conversion happens before the spiral
reaches the boundary. In the presence of electrophoretic driving with
$\eleamp=0.03$, we find that the minimal shock strength required for
conversion is $\shockamp\conv\approx0.1489$ (the case shown in the figure),
which is slightly smaller than that $\shockamp\conv\approx0.1789$ required for
conversion in the absence of applied field ($\eleamp=0$).


\subsection{Verfication of predictions and effects of discretization}
\seclabel{coef}

\sglfigure{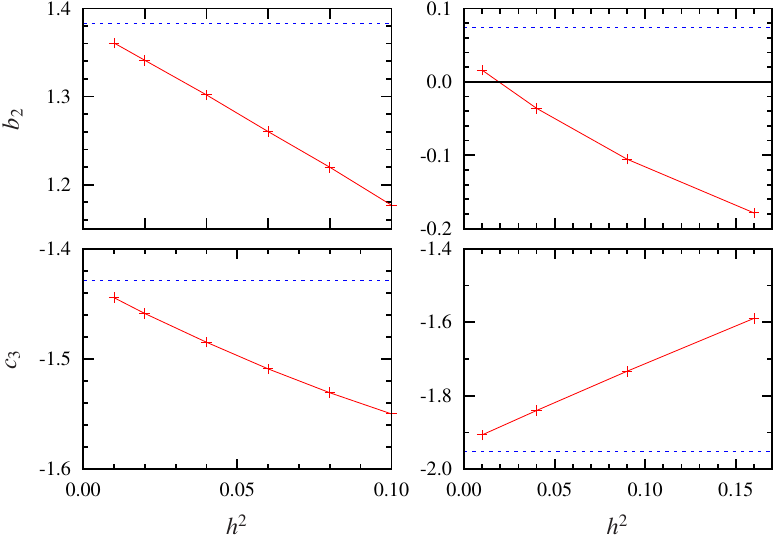}{
  (Color online) Numerical convergence of $\btwo$ and
  $\cthree$ from numerical simulations of electrophoretic
    drift. 
  The fast vortex is simulated over regular grid spacings $\dr=\dz=\h$ as 
  indicated.
  Left column: $\parbet=0.71$, a typical case away from $\btwo = 0$.  
  Both coefficients converge quadratically in $\h$ to the values predicted by
  RFs (blue dashed lines). 
  Right column: $\parbet=0.722$, a case close to $\btwo = 0$ where filament
  tension is small.
  $\cthree$ converges quadratically while the convergence of $\btwo$ is less
  clear. Importantly, except at very high resolution the tension in the
  simulation is of the opposite sign from the asymptotic result. 
}{conv}

Before proceeding to 3D simulations, we compared the predictions
for the filament tension $\btwo$ and binormal drift coefficient $\cthree$ 
obtained via RFs in \secn{predict}, with 
direct 2D computer simulations of axisymmetric scroll rings as discussed in
\secn{rings}. 
From the motion of the axisymmetric filament
in the radial (normal) and vertical (binormal) directions we obtain the
coefficients $\btwo$ and $\cthree$. In 2D, it is possible
to perform simulations to very high spatial resolution with reasonable cost
and hence we are also able to study systematically the effects of 
numerical resolution.

Simulations of electrophoretic drift at $\eleamp=0.01$
have been conducted for the fast vortex over a variety of regular
grids with spacings $\dr=\dz=\h$. Results are shown in \fig{conv} for two
values of $\parbet$: $\parbet=0.71$, a typical case away from where $\btwo$
crosses zero in \fig{fig1}, and $\parbet=0.722$, a case very near the zero
crossing of $\btwo$.  In the typical case, left column of \fig{conv}, both
$\btwo$ and $\cthree$ converge extremely well as the grid size goes to zero
to the predicted values given by Eq.~\eq{b2_coeff}. The convergence is
quadratic, \ie\ linear in $\h^2$, as is consistent with the second-order accuracy
of the numerical simulation.

The case shown in the right column of \fig{conv}, where the predicted value of
$\btwo$ is small, is more problematic. Again, $\cthree$ converge extremely
well, and with the expected form, to the predicted result. The convergence of
the tension $\btwo$ is less satisfying, both in the form of the convergence
and in the asymptotic value. The exact reasons for this are outside the focus
of the current study. The important point is the sign change of $\btwo$ seen
in the right column of \fig{conv} due to finite resolution. This means that a
3D numerical simulation performed at even high resolution, \eg\ $\h=0.2$ will be
qualitatively different from a fully resolved simulation at these parameter
values, since the simulated filament tension will be negative whereas the
fully resolved tension is positive.

The conclusion is that the filament tension and binormal drift predicted from
RFs and plotted in \fig{fig1} are borne out by direct numerical
simulations. Nevertheless, 3D simulations should only be performed for the
model parameters such that the discretization does not qualitatively affect
the simulations by artificially changing the sign of the filament tension. For
the parameters we consider, this means to the left ($\parbet \le 0.71$) or the
right ($\parbet \ge 0.73$) sides of zero crossing of $\btwo$ in \fig{fig1}.



\section{Three-dimensional simulations}
\seclabel{3D}

We have seen in \secn{2D} that it is possible convert between fast and
slow vortices in 2D by a variety of mechanisms, such as interactions with
boundaries and shocks. These effects will also exist in 3D, but in 3D the
additional effects arising due to filament curvature and filament tension can
be hugely important. This is especially so in the region where, according to
predictions of \secn{predict}, the alternative
vortices have opposite signs of filament tension. 

Simulations have been
performed with a version of EZSCROLL~\cite{Dowle-etal-1997,ezscroll}
modified for FHN
kinetics.  The simulations use forward Euler timestepping on a uniform
Cartesian grid on cuboid domains of varying size with non-flux boundary
conditions, nineteen-point approximation of the Laplacian, space
discretization step $\dx=\dy=\dz=\h=1/3$ and time discretiation step
$\dt=3/80$.  Numerical determination of the instant position of scroll
filaments is technically challenging, and we use an easy substitute: the
instant phase singularity lines, defined as the intersections of isolines
$u(x,y)=u_*$ and $v(x,y)=v_*$.  Such singularity lines correspond to spiral
tips in 2D, and like a spiral tip rotates around the current rotation center,
so a singularity line rotates around the current filament. We also assume that
inasmuch as the twist of scroll waves is insignificant, then the curvature of
the singularity line is close to the curvature of the filament, so we take the
former as an approximation of the latter.  As with 2D simulations in EZSPIRAL,
we use $(u_*,v_*)=(0,0)$ for the FitzHugh-Nagumo model, with some exceptions,
described later.  Initial conditions for alternative scrolls are obtained by
first computing alternative spirals on a polar grid, using methods described
in Ref.~\onlinecite{Biktasheva-etal-2009}, then converting these solutions to
Cartesian coordinates and ``stacking'' spirals on top of one other to generate
3D scroll waves. 

It is more difficult in 3D than in 2D to use core size and visual inspection
to distinguish between fast and slow vortices. However, point records of the
action potentials, $u(t)$ at some fixed location, may clearly distinguish the
two types of vortices. Recall \fig{intro_own}. Slow vortices have extra maxima
in the tails of their action potentials, the DADs, while fast vortices do not,
no DADs.  This method of distinguishing between vortices assumes that the
whole solution consists of only slow or only fast vortices. A more
sophisticated method free from this assumption is described later.


\subsection{Alternative vortices with opposite signs of filament tension} 

\sglfigure{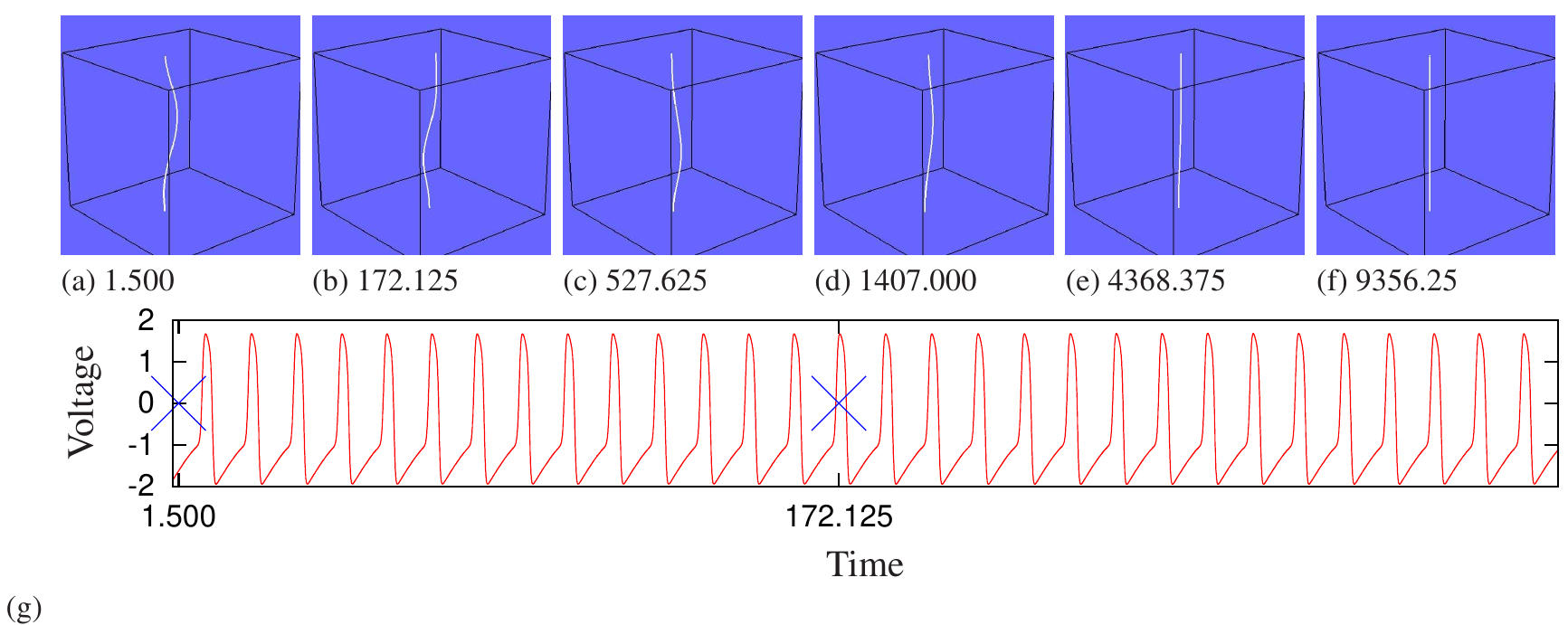}{
  (Color online)
  Evolution of a fast helical scroll
  with positive filament tension at $\parbet=0.71$. 
  The domain has Neumann boundary conditions. 
  (a-f) The
  filament straightens up (the numbers under the snapshots show corresponding
  time in timeunits); (g) action potentials with blue crosses marking the
  times at which two of the snapshots are taken.  
  Singularity lines are used as approximations to the scroll filament.
}{3D_b0.71_inc_nbc}

\sglfigure{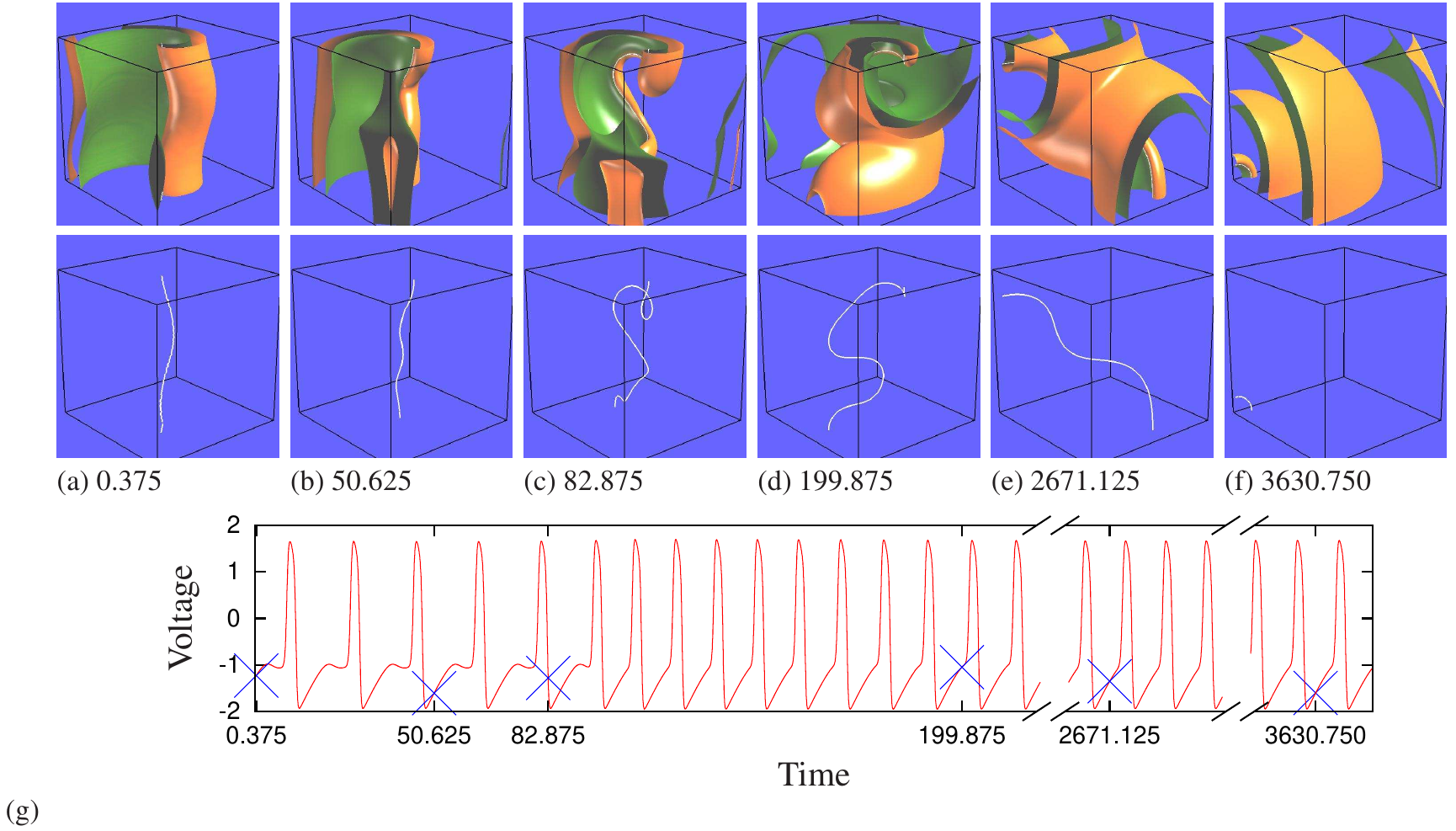}{
  (Color online)
  Evolution of a slow helical scroll
  with negative filament tension at $\parbet=0.71$.
  Same domain and parameters as 
  for the fast vortex in \fig{3D_b0.71_inc_nbc}. 
  (a-f) (Top) isosurfaces of the $u$-field; (middle)
  filament only (as in \fig{3D_b0.71_inc_nbc}). 
  (g) action potentials with blue crosses marking the
  times at which the snapshots are taken. 
  The helix initially expands as expected for negative tension, but then
  undergoes conversion and contracts. 
}{3D_b0.71_dec_nbc}

The first case we consider is at $\parbet = 0.71$, where the fast and slow
vortices have opposite signs of filament tension.  
Simulations have been conducted in a box of size $50\times50\times50$\:s.u.,
with Neumann boundary conditions (NBCs) on all sides.  Simulations are started
from alternative vortices with filaments in the form of helices constructed by
layering 2D solutions such as to create a helical scroll with one turn from
bottom to top. The radius of the helix is 2\:s.u.
We expect that the filament of the fast vortex with positive filament tension
will straighten up while slow vortex with negative filament tension will
expand, possibly breaking up and developing into turbulence.

\Fig{3D_b0.71_inc_nbc} shows the evolution of the fast vortex with
positive filament tension. As expected, the filament straightens up,
see \FIG{3D_b0.71_inc_nbc}{a-f}. The record of corresponding action
potential, \FIG{3D_b0.71_inc_nbc}{g}, shows that there is no
conversion into the slow alternative vortex: no 
DADs appeared.

\Fig{3D_b0.71_dec_nbc} shows the evolution of the slow vortex with negative
filament tension. First, the filament expands as expected. However, after five
full rotations, the vortex spontaneously changes its period and converts into
its fast counterpart with positive filament tension, and we see DADs
disappearing at this moment in \FIG{3D_b0.71_dec_nbc}{g}. The fast, positive
filament tension vortex subsequently contracts, see
\FIG{3D_b0.71_dec_nbc}{d-f}, and disappears.


\subsection{Conversion due to curvature}

To identify the cause for the spontaneous conversion in the simulations just
discussed, we have repeated the simulation with the slow helical vortex 
in different domain sizes with periodic boundary
conditions (PBCs) on the top and bottom of the domain, and NBCs elsewhere.

In this series of simulations, we start from a helical filament with
one full turn from bottom to top.
We monitor the mean radius of the
projected filament onto the base of the box. Initially the projection is a
circle. The initial helical filament expands.  As this expansion is 
inherently unstable, the projections of the filament onto the base
very soon cease to be circular. See
\figs{0.71_proj_50x50x50}{0.71_proj_70x70x50}. 

\Fig{radii}{(a)}
shows how the mean radius of a vortex filament's projection changes every
period in the boxes $50\times50\times50$\:s.u.,
$50\times50\times100$\:s.u. and $70\times70\times50$\:s.u. The vertical lines
show the time at which conversion happens in each box, as indicated by the
morphology of the point records. It can be seen that in the boxes
$50\times50\times50$\:s.u. and $70\times70\times50$\:s.u., both of height
$L_z=50$\:s.u., conversion happens at approximately the same time, regardless
of the width of the box. Whereas in the box $50\times50\times100$\:s.u. with
the height $100$s.u. conversion happens much later than in the other two
cases.  Comparing the simulations in boxes $50\times50\times50$\:s.u. and
$50\times50\times100$\:s.u., one can conclude that interaction with the
boundary is probably not the main factor in the conversion, as it takes
significantly different times in the boxes of the same width, and similar times
in boxes of different widths.

Critical curvature is a more plausible cause for conversion.  On
qualitative level, larger $L_z$ means smaller initial curvature and,
by Eq.~\eq{fil-motion}, slower evolution. This is indeed what happens: the helix
with $L_z=100$ expands more slowly than those with $L_z=50$. See 
\fig{radii}(a). Thus it will take longer for the
filament curvature to reach any particular value, \eg\ its critical value.

\sglfigure{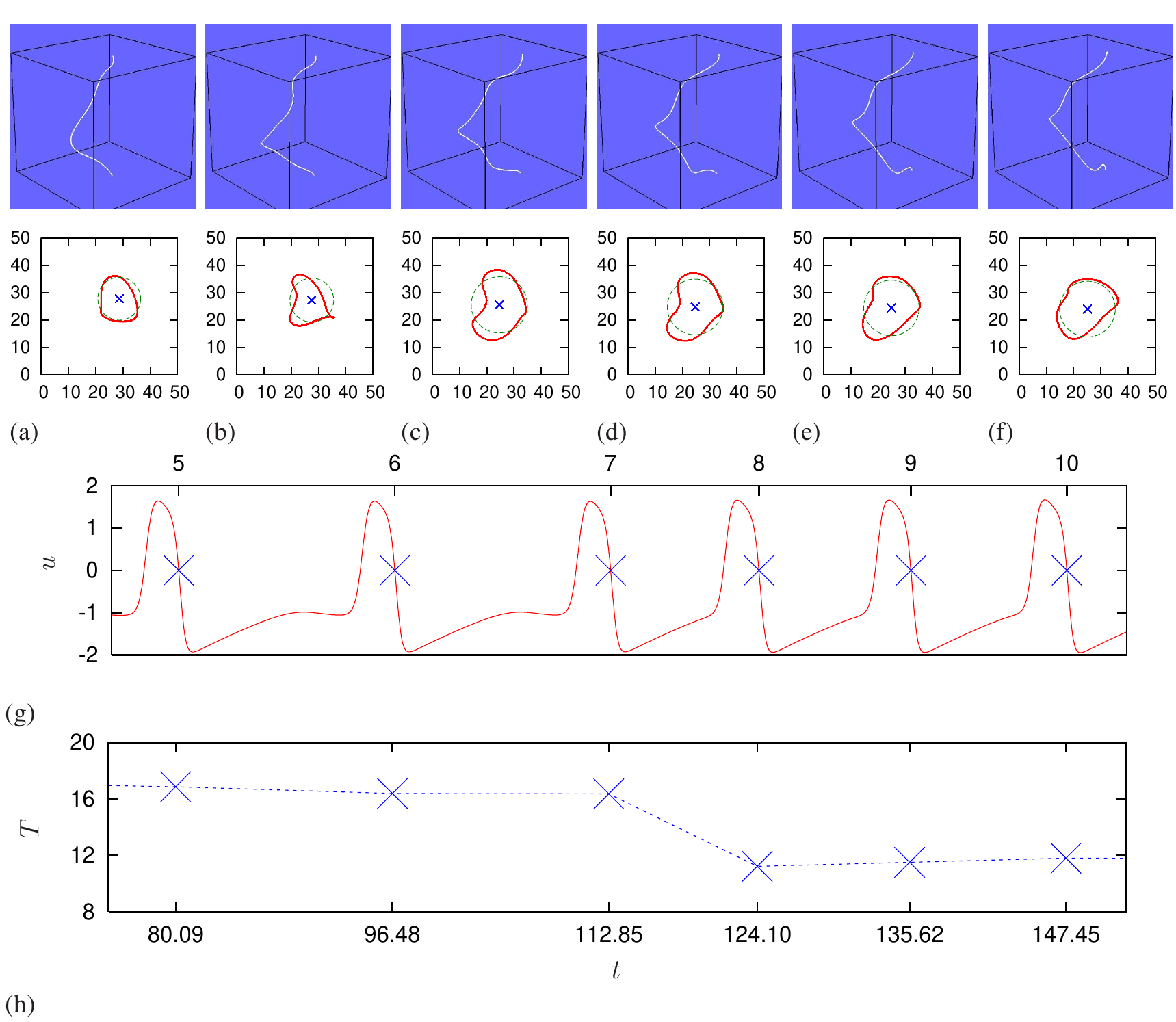}{
   (Color online) Conversion of
  a slow helical vortex into a fast one at $\parbet=0.71$. Domain size is
  $50\times50\times50$\:s.u. with PBC on top and bottom and NBC elsewhere.
  (a-f) evolution of the filament (top row) and projections of the
  filament onto the base of the box (second row) shown in red, with
  the barycenter shown as a blue cross and the mean radius shown as
  the green circle. (g) action potentials ($u$-field against time $t$),
  blue crosses correspond to the times at which the snapshots in (a-f)
  are taken, and the period number is shown on the upper
  axis. (h) period $T$ plotted against time $t$.
}{0.71_proj_50x50x50}

\sglfigure{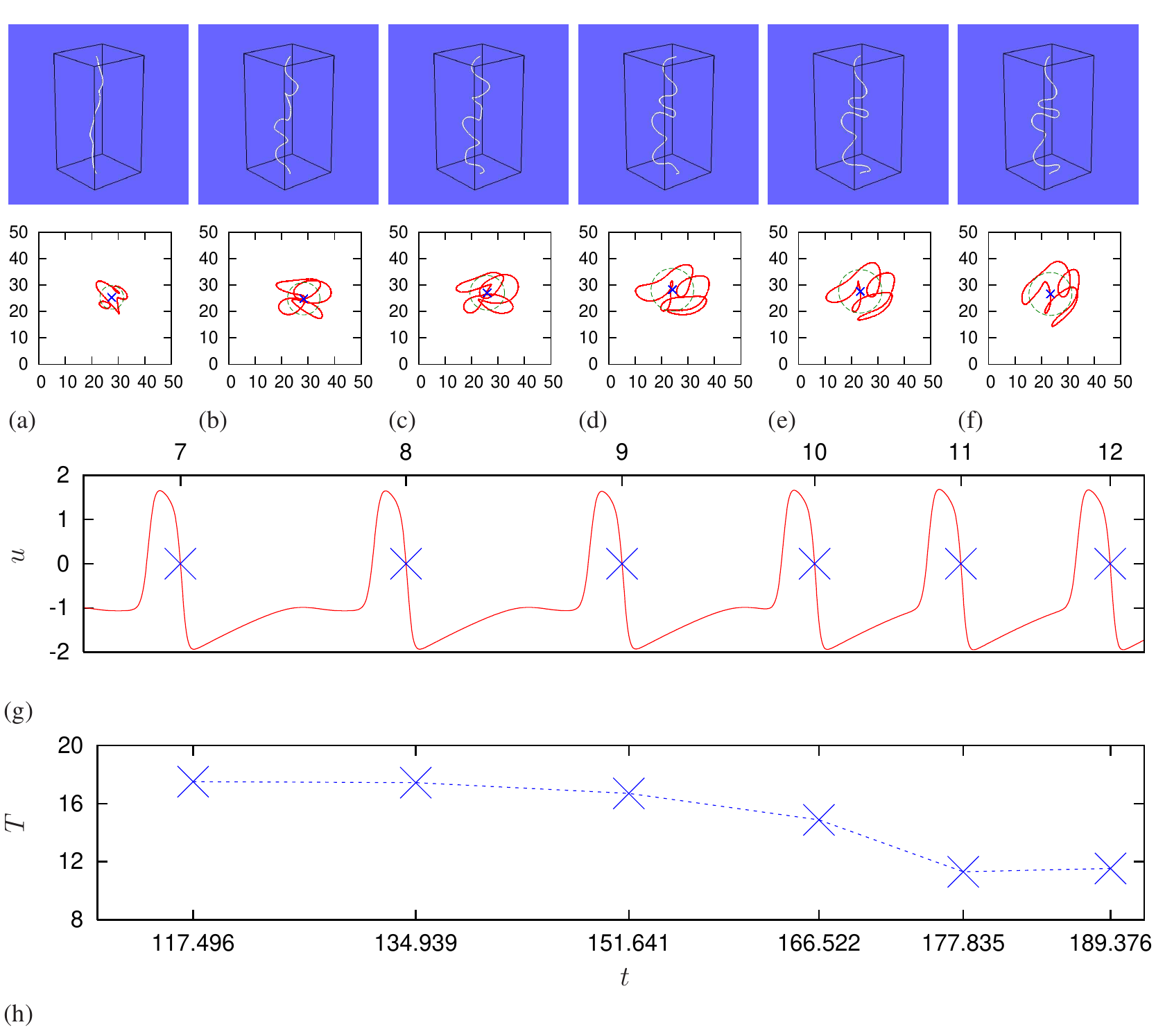}{
  (Color online)
  Same as in \fig{0.71_proj_50x50x50} except with a box of size 
  $50\times50\times100$\:s.u.
}{0.71_proj_50x50x100}

\sglfigure{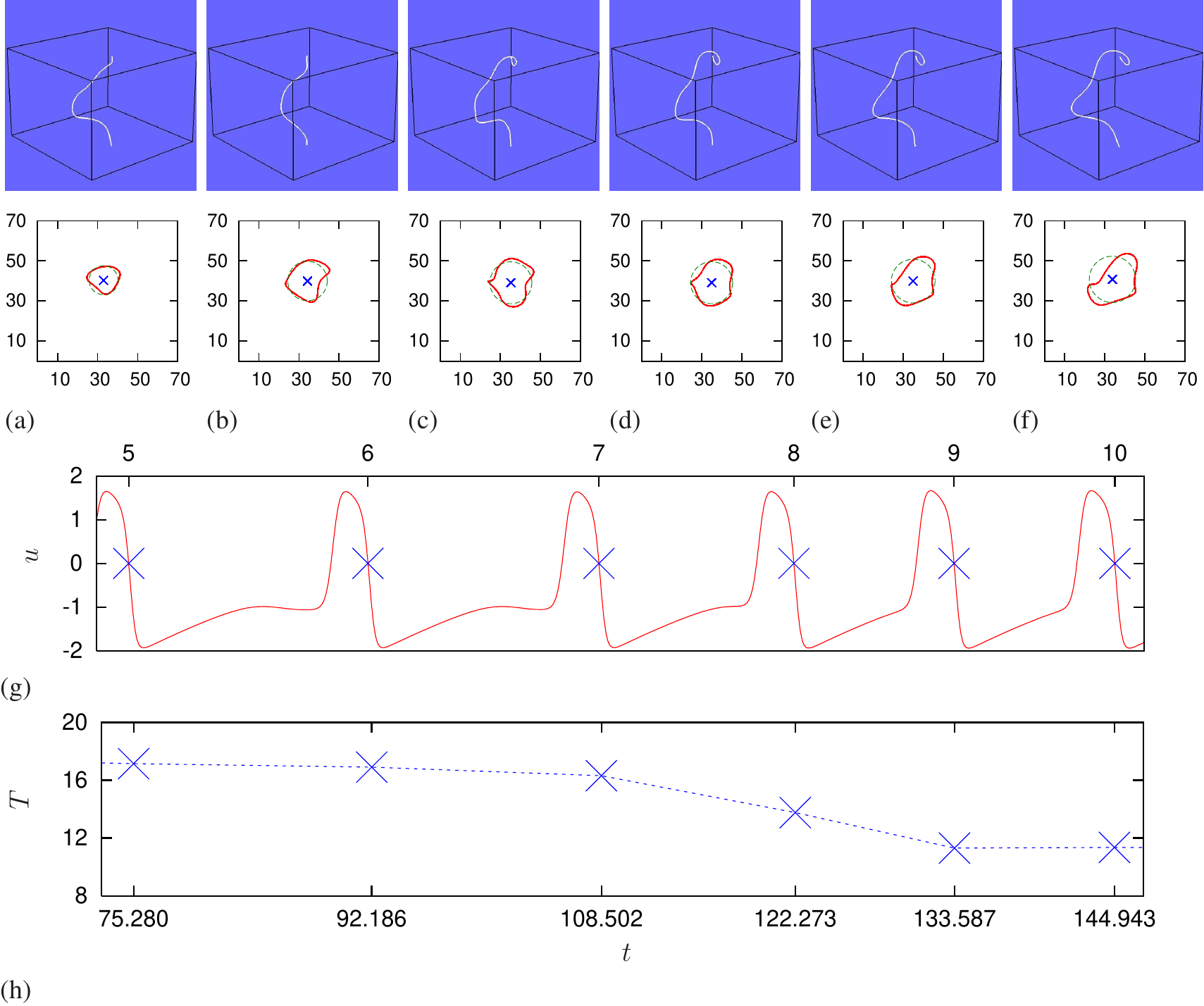}{
  (Color online)
  Same as in \fig{0.71_proj_50x50x50} except with a box of size 
  $70\times70\times50$\:s.u.
}{0.71_proj_70x70x50}

\sglfigure{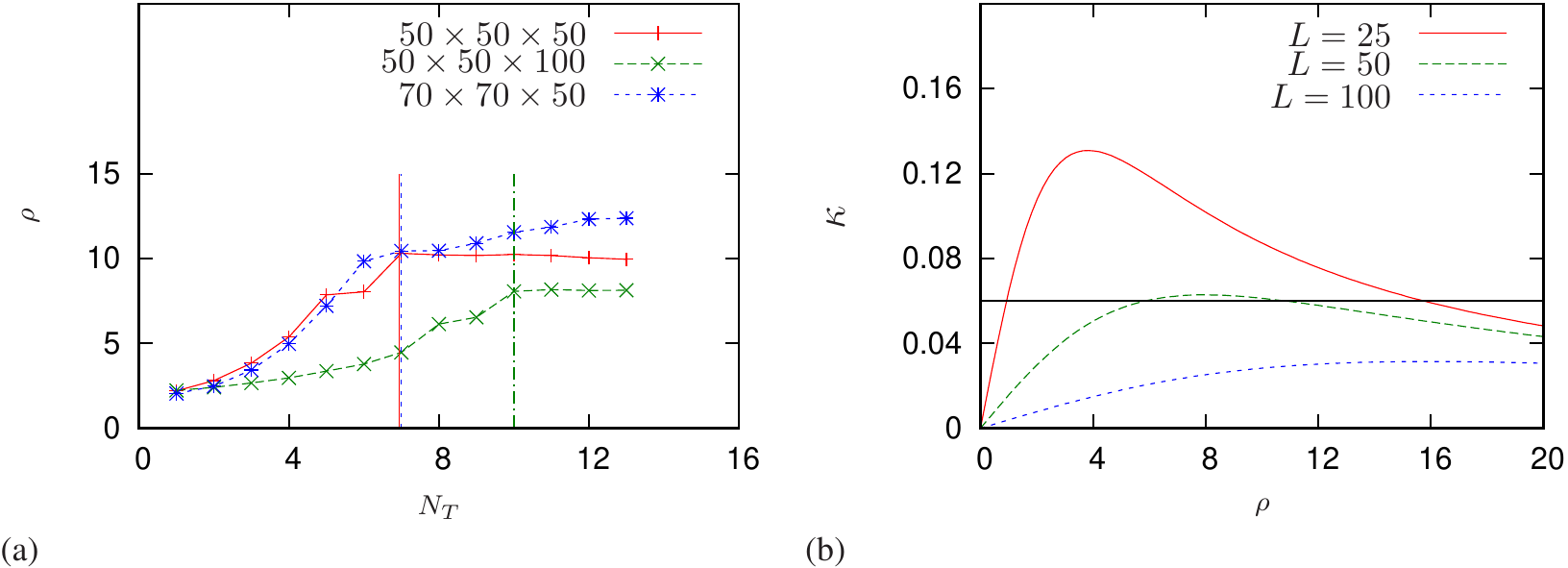}{
  (Color online)
  Curvature evolution. (a) The mean radius, $\rho$, of the
  $(x,y)$ projection of the filament, against the period number,
  $N_T$. The vertical lines separate the slow region (to the left of
  the line) from the fast region (to the right). (b) Dependence of
  the curvature, $\curv$, on the radius of the projection, $\rho$,
  using Eq.~\eq{curv_theory} for an ideal helix. 
  The horizontal line is the critical curvature, $\curv\crit$.
}{radii}

To quantify this argument, consider an ideal helix whose
curvature $\curv$ depends on the radius of its projection $\rho$ as
\begin{equation}
\curv = \frac{\rho}{\rho^2 +\left(L_z/2\pi\right)^2} \, ,
\label{curv_theory}
\end{equation}
where $L_z$ is the height of the helix making one full turn. 
\Fig{radii}(b) shows how the curvature $\curv$ of an ideal helix
changes with $\rho$, for the selected values of $L_z$. The important feature
is that 
the curvature of a helix increases with $\rho$ up to a certain maximum which
depends on $L_z$.
The larger $L_z$, the slower the growth of $\curv$ with $\rho$ and the smaller
the maximum.  For comparison, we also show on \fig{radii}(b) the curvature
$\curv\crit$, which causes immediate conversion. This is known from the 2D
simulation shown in \FIG{electro_shock_amps}{c} where $\eleamp=0.06$ causes
immediate conversion. Again using the formal equivalence between
electrophoresis and curvature (\secn{rings}), we have
$\curv\crit=0.06$.

These graphs imply that for an ideal vortex, the conversion due to curvature
will happen later for larger $L_z$, which is in qualitative agreement with the
observations shown in
\fig{radii}(a). However, there is no quantitative agreement. This is likely
because the shape of the filaments very soon deviates from an ideal helix.

We have explicitly calculated the curvature of the filaments using numerical
differentiation with Tikhonov regularization.  The details of the procedure
can be found in Appendix~A.
The results of this analysis are presented in
\FIG{775dbl}{a} and \FIG{5510dbl}{a}, with the local curvature shown against $z$
coordinate on the filament, for selected values of the scroll period numbers
$N_T$. We see that the curvature grows very non-uniformly, both along the 
filament and in time.
 
To detect the conversion visually, so as to be more precise about time and
location of the conversion, we have analyzed the fine structure of the scroll
cores, using a modified version of the instant phase singularities:
$u_*=-1.04$, $v_*=-0.656$. The idea is that the point $(u_*,v_*)$ is within
the small loop of the phase trajectory shown in
\fig{intro_own}(c), 
and existence of such a loop is the signature of slow scrolls.  Using this
definition, slow scrolls are characterized by a double singularity line,
whereas the fast scrolls have a single singularity line.
See \FIG{775dbl}{b-d} and \FIG{5510dbl}{b-d}.  One can see that the conversion
is indicated by the departing of the secondary singularity from the
main one following the period in which the filament curvature (obtained for
the standard, ``robust'' singularity $u_*=v_*=0$) has exceeded threshold on a
substantial continuous interval. We see this
as a confirmation that curvature of the filament is the likely cause of the
conversion in these simulations.

\dblfigure{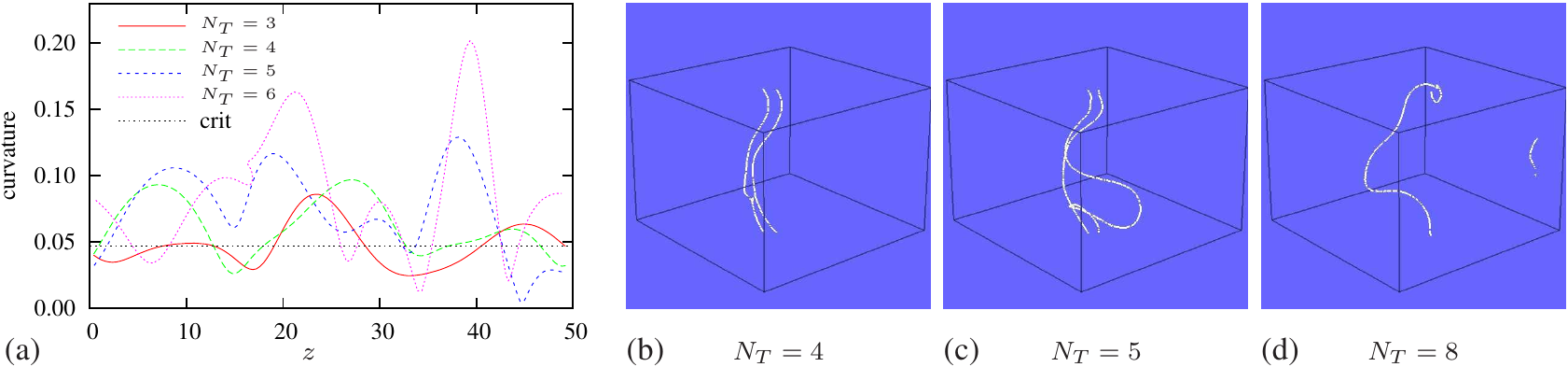}{
  (Color online)
  Curvature and conversion in the $70\times70\times50$\:s.u. domain 
  corresponding
  to the simulation shown in \fig{0.71_proj_70x70x50}. 
  (a) The curvature against $z$
  coordinate on the filament, for selected values of the scroll period
  numbers $N_T$. (b,c,d): the double-singularity visualization of the
  scrolls (b) immediately before, (c) in the very beginning, and (d) after
  completion of the conversion.  
}{775dbl}

\dblfigure{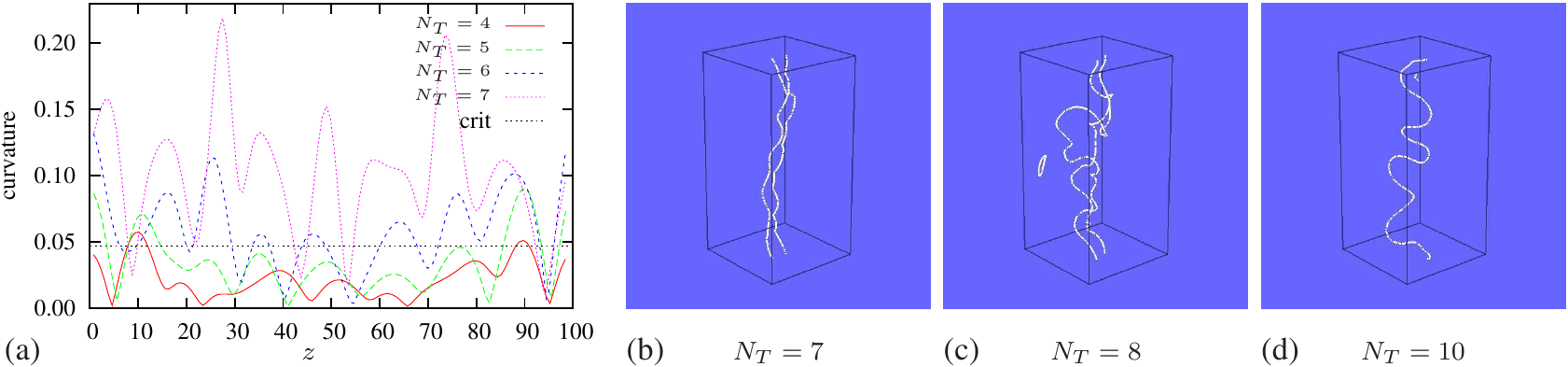}{
  (Color online)
  Same as in \fig{775dbl}, except for the 
  $50\times50\times50$\:s.u. simulation shown in
  \fig{0.71_proj_50x50x100}.
}{5510dbl}


\subsection{Dynamics of alternative ring vortices}

\sglfigure{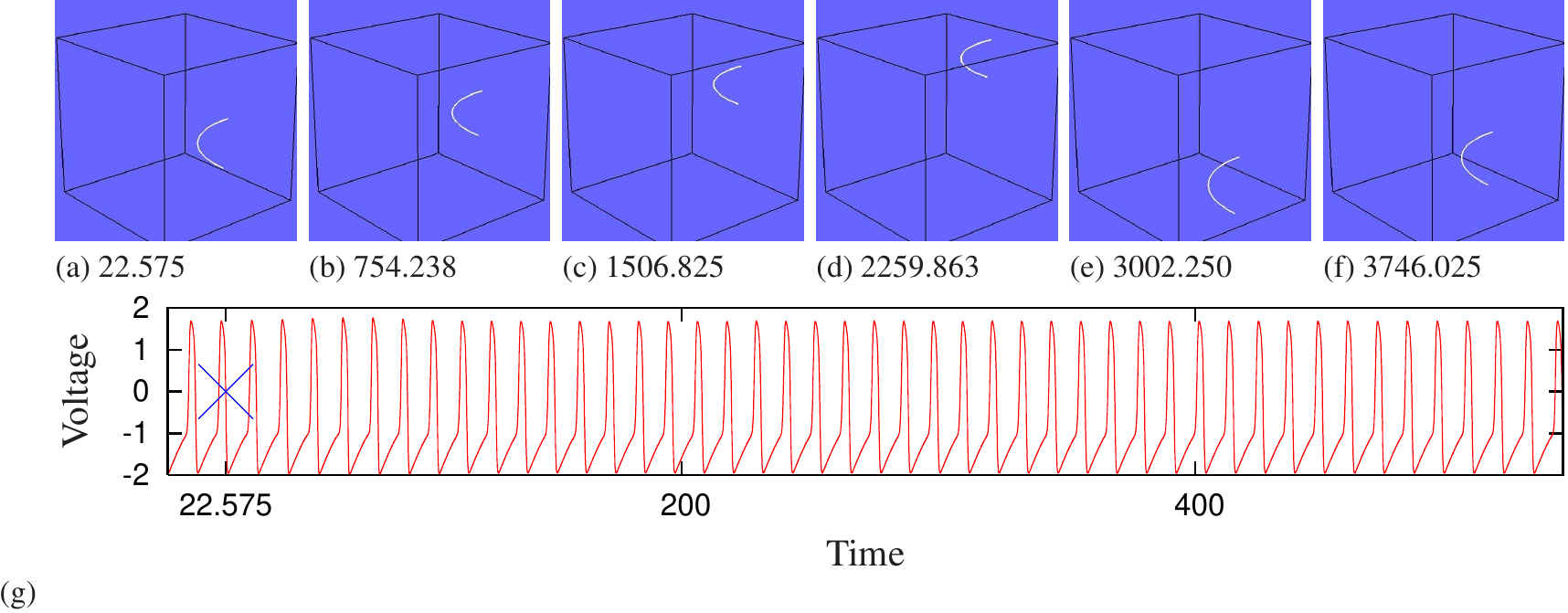}{
   (Color online) Evolution of a fast scroll ring with positive tension at
   $\parbet=0.71$.  One quarter of the ring is simulated in a domain with
   Neumann boundary conditions on four sides and periodic boundary conditions
   one the top and bottom. Visualizations and time series are as in
   previous figures. 
}{3D_b_0.71_inc_quarter_pbc}

We have also considered the dynamics of alternative scroll rings with
filament tension of opposite sign at $\parbet=0.71$.  One expects that, at
least initially, a ring with positive tension will drift and contract, while a
ring with negative filament tension will drift and expand. We are interested
in the long-term dynamics of such rings.

A quarter of a scroll ring with positive filament tension is initiated in a
cubical box of size $50\times50\times50$\:s.u. Two cases have been considered.
In one the box has Neumann boundary conditions on all sides (not shown) and in
the other the box has Neumann boundary conditions on four sides and periodic
boundary conditions on the top and bottom (case shown
in \fig{3D_b_0.71_inc_quarter_pbc}). The initial radius of the ring is
$\approx25$\:s.u. and meets two adjacent sides at right angles as seen
in \FIG{3D_b_0.71_inc_quarter_pbc}{a}.

The scroll drifted upwards, remaining essentially flat. In the case with
Neumann boundary conditions on all sides, the ring approaches the upper
boundary where it shrinks and eventually disappears in a top corner of the
box. With periodic boundary condition, as seen in
\fig{3D_b_0.71_inc_quarter_pbc}, instead of collapsing, the ring
drifts upwards continuously and reaches an asymptotic constant radius. 
In a way, this perpetual
movement is similar to what was observed in 2D simulations of electrophoretic
drift with periodical boundary conditions, \FIG{electro_nbc_0.71}{d-f}.  

The existence of the stable vortex rings in excitable media was first
reported by Winfree~\cite{Winfree-1994}. 
The upward component of the ring's velocity at the chosen set of
model parameters is in agreement with the asymptotic theory
which gives $\cthree<0$ for $\parbet=0.71$. 
More importantly, it
seems that the positive filament tension, $\btwo>0$, on its own may be not
enough for a ring to collapse. It seems that, in the absence of other 
perturbations, the
positive filament tension  shrinks a ring vortex just to a stable
minimum radius not equal to zero, while  the internal interaction of
parts of the vortex prevents it from complete collapse.
Note that in simulations of vortex rings with positive filament tension, with
both Neumann and periodic boundary conditions, there no conversion of the fast
vortex into its slow counterpart has been observed.

\sglfigure{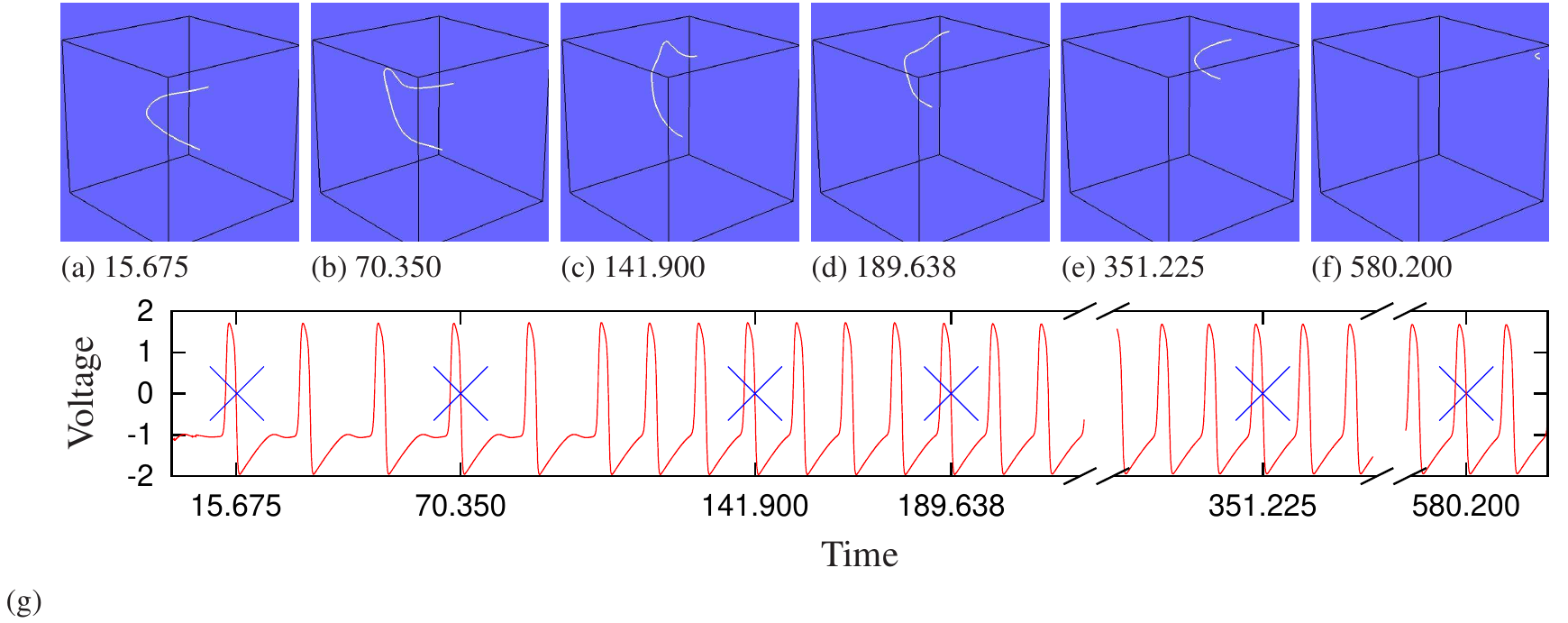}{
   (Color online) Evolution of a slow scroll ring with negative tension at
   $\parbet=0.71$.  The simulation is initiate with a quarter of the ring 
   a domain with Neumann boundary conditions on all sides.
   Visualizations and time series are as in previous figures. 
}{3D_b0.71_dec_quarter_nbc}

Now consider the evolution of the slow vortex ring with negative filament
tension shown in \fig{3D_b0.71_dec_quarter_nbc}.  The model parameters are the
same as in the previous simulation although here we show the case with Neumann
boundary conditions on all sides of the domain.  In accordance with the
negative filament tension, the ring initially expands
[\FIG{3D_b0.71_dec_quarter_nbc}{a-b}] and becomes non-planar.  After five
rotations the vortex converts into its fast counterpart (with positive
tension), as seen in the action potential recordings
in \FIG{3D_b0.71_dec_quarter_nbc}{g}.  From this point the vortex propagates
to the top boundary where it contracts and disappears in the top
corner, \FIG{3D_b0.71_dec_quarter_nbc}{c-f}.  

In case of periodic boundary
conditions (not shown) the scenario is: expansion due to negative tension
followed by conversion to the fast vortex with positive filament tension,
followed by contraction to a stable ring with a nonzero radius, similar
to \fig{3D_b_0.71_inc_quarter_pbc}.
The same type of scenario was first
reported by Sutcliffe and
Winfree~\cite{Sutcliffe-Winfree-2003} who explained it by the twist
of the filament. 
Here we have demonstrated that the switching from expansion to
contraction could be due to simple switching from a vortex with
negative filament tension to its counterpart with positive filament
tension.


\subsection{Alternative vortices with negative filament tension}

At $\parbet=0.73$, the alternative solutions both have negative filament
tension, as can be seen \fig{fig1}. Still, only the slower vortex
displays DADs in its action potentials. As we know from the 2D studies
[\fig{b2}], at
these model parameters conversion is possible both ways, from slow to fast and
from fast to slow vortex.

First we consider the evolution of a helical vortex initiated from the fast
spiral shown in \fig{3D_b0.73_inc}.  Simulations are in a box of size
$50\times50\times50$\;s.u., with Neumann boundary conditions on all sides.
The filament initially expands due to negative tension, but quickly converts
to the slow scroll after just two rotations. See the DADs appeared
in \FIG{3D_b0.73_inc}{g}. The slower vortex continues to expand and then
converts back to the fast vortex after 14 periods. Since both alternative
vortices have negative tension, eventually full fibrillation with multiple
filaments develops as is seen \FIG{3D_b0.73_inc}{c}. There are no further
spontaneous conversions, and the fast vortex with negative filament tension
continues to evolve. However, we see a gradual reduction in the number of
filaments until only a single tiny piece persists in the bottom right corner
of the box.

\sglfigure{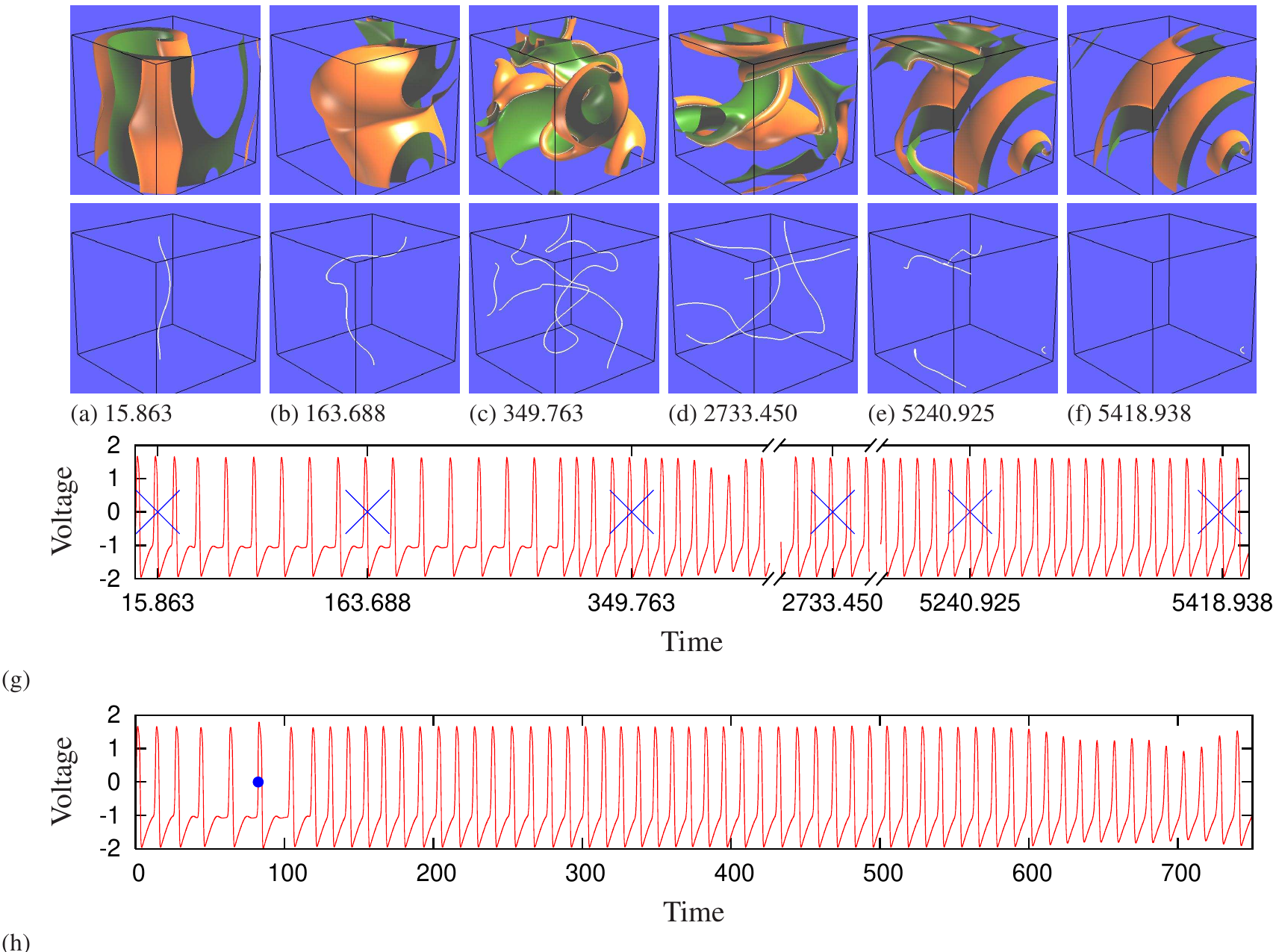}{
  (Color online)
   Evolution when alternative scrolls both have negative filament tension,
   $\parbet=0.73$. Simulations start with a fast helical scroll. The boundary
   conditions are Neumann on all boundaries. 
   (a--f) (Top row) isosurfaces of the $u$-field; (second row)
   filaments only. 
   (g) Action potentials with blue crosses marking the times at which
   the snapshots are taken. (h) Action potentials for a simulation
   with the same initial and boundary conditions, but with a shock
   stimulus applied at the instant marked with a blue dot). 
}{3D_b0.73_inc}

Thus far, we have observed only spontaneous conversion between alternative
vortices in 3D. It is important for cardiological applications to verify
whether 3D vortices can be converted by a shock. Having the prolonged period
of slow vortex evolution in the simulation shown in \FIG{3D_b0.73_inc}{g}, we
have repeated the simulation and this time applying a uniform shock,
Eq.~\eq{shock}, to the slow vortex.  The action potential time series, but not
the visualization, are shown in \FIG{3D_b0.73_inc}{h}.  A shock of amplitude
$\shockamp=0.8$ is applied just after the third slow vortex period.
The shock successfully converts the slow vortex into its fast counterpart. The
rest of the evolution of the fast scroll was similar to
\FIG{3D_b0.73_inc}{a-f}: development of the full fibrillation followed
by gradual reduction of the number of filaments in the box to just a single
tiny persistent piece in the top right corner of the box, but in approximately
half of the time taken without the shock.




\section{Discussion}
\setcounter{paragraph}{0}

In this study we have investigated dynamics of alternative
spiral and scroll waves in the FitzHugh-Nagumo model, 
in a parameter region of bistability between distinct alternative vortices.
Some of the features have been noted already
by Winfree~\cite{Winfree-1991a} who discovered such alternative
solutions. Here we have extended those studies and related them to the
asymptotic theory of spiral and scroll wave dynamics. The most
important features are:
\begin{enumerate}
\item\label{alt-spirals} There is a parameter region in FHN model
  where stable alternative spiral wave solutions, with different rotation
  frequencies, exist. Faster spiral waves have the normal morphology of the
  ``action potential'' whereas the slower spiral waves show ``delayed
  after-depolarization'' morphology.
\item\label{shock-conversion} The alternative vortices can convert one
  into the other as a result of a variety of perturbations, such as 
  a homogeneous pulse 
  stimulus or interaction with a boundary.
  There are distinct parametric regions where this transition may
  occur in one direction only, either from slow to fast vortex or from
  fast to slow, as well as the regions where the transition can 
  occur
  in both directions. Conversion of a single slow vortex into
  multiple fast vortices is also possible. 
\item\label{spontaneous-conversion} The conversion effects of spirals
  can also be observed for scroll waves. In addition, scroll waves can
  demonstrate spontaneous conversion apparently related to the
  filament curvature. 
\item\label{alternative-tension} In accordance with the predictions of
  the asymptotic theory, the alternative scroll waves have
  significantly different filament tensions. In particular, the filament tension of the faster
  scroll waves changes sign in the
  parametric interval considered, whereas the slow scroll waves have
  negative filament tension in this interval. Therefore, there exists
  a parameter region where the fast vortex with positive filament
  tension has a slow counterpart with negative filament tension, and a
  parameter region where both alternative vortices have negative
  filament tension.
  The scrolls with positive filament
  tension have tendency to contract, or straighten up if the filament connects
  opposite surfaces of the box. The vortex rings with positive
  filament tension might have a stable radius not equal to zero, and
  collapse only due to an additional perturbation, e.g. hitting the
  boundary perpendicular to the axis of the ring.
\end{enumerate}

The scrolls with negative filament tension have the tendency to
lengthen and multiply, which can lead to ``scroll wave turbulence'',
phenomenologically similar to certain stages of cardiac fibrillation.
However, interaction of different scroll filaments with each other
and/or with the boundary may lead to the stabilization of the scrolls
notwithstanding the effects of the filament tension,
as was observed previously (e.g. Ref.~\onlinecite{Biktashev-1989}
  p.~139) and in simulations presented here.
Therefore, in three dimensional experiments and simulations
the stabilized scrolls with negative
filament tension may behave identically to the scrolls with the
positive filament tension, and the only way to distinguish between
them is to compute/measure their response functions.

The conversion processes summarized in points \ref{shock-conversion} and
\ref{spontaneous-conversion} above 
are essentially threshold effects and so are
not asymptotic in nature. However, there are examples where
(non-asymptotic) analytical approaches have been successful in
describing threshold phenomena,~\cite{Idris-Biktashev-2008}. Following
the idea of Ref.~\onlinecite{Idris-Biktashev-2008}, it may be possible
to describe conversion of
spiral waves in response to external stimuli using center-stable space
of the unstable spiral wave solutions, which presumably separates (in
the functional space) the stable alternative spirals observed in
simulations.

Assuming that the above features are present in other models, including more
physiologically realistic ones,
one may conjecture the following scenarios,
which may be relevant to cardiac electrophysiology.

\emph{Shock-induced conversion of fibrillation to tachycardia.}
Suppose the cardiac tissue fibrillates due to ``scroll turbulence''
mechanism, underlied by slow scrolls. Such slow scrolls are likely to
convert to fast scrolls either due to curvature or to interaction with
each other or with boundaries. However this may take a long time.  An
electric shock which is too weak to instantly defibrillate, may still
be enough to convert from slow to fast scrolls. If the fast scrolls
have positive filament tension, this may lead to ``delayed
defibrillation'', when the fibrillation stops via collapse of all
scrolls, but only many cycles after the shock, or to ``tachycardia''
when the fast scrolls would stabilize either by connecting opposite
surfaces of the tissue (say transmural filaments) or by attaching to
localized inhomogeneities or anatomical features. 

If the corresponding fast scrolls have negative filament tension, they
might stabilize due to interaction with each other and with the
boundary, leading to tachycardia indistinguishable from the one with
positive filament tension and without pinning to anatomical obstacles.

\emph{Intermittent fibrillation.}
If both alternative scrolls have negative filament tension and can be
converted equally one into another, then 
a relatively weak
external shock applied to the stabilized fast scroll can convert it
back into the slow one, which will initiate another, may be prolonged
fibrillation period before eventual conversion and stabilization back
into tachycardia.

FitzHugh-Nagumo model is admittedly very far from realistic cardiac
models. However, delayed after-depolarization is indeed known in
electrophysiology, and known to have arrhythmogenic
effect~\cite{Wit-Rosen-1983,Wu-etal-2004}. 
The presence of DADs will add to the period of
the vortices, and the tendency of vortices with longer periods to have
negative filament tension is universal\cite{Brazhnik-etal-1987} and
not restricted to the FHN model. The coexistence of vortices with
different periods and different filament tensions is key to the
effects we have described. Hence we believe that the basic
phenomenology involved in our results is not restricted to FHN and may
be observed in any excitable media with DADs, and so is likely to be
relevant to cardiac electrophysiology.
This capacity of DAD is distinct from its well-known role as a mechanism
of arrhythmia initiation. 


\section*{Acknowledgement}
This study has been supported in part by EPSRC grants EP/D074789/1 and
EP/D074746/1. DB also acknowledges support from the Leverhulme Trust and
the Royal Society.

%

\end{document}